\documentclass{osa-article}
\journal{oe}

\begin{document}
\title{Cross talk compensation in multimode continuous-variable entanglement distribution}

\author{Olena Kovalenko,\authormark{1,*} Vladyslav C. Usenko,\authormark{1} and Radim Filip\authormark{1}}

\address{\authormark{1}Department of Optics, Palack\'y University, 17. listopadu 12, 771 46 Olomouc, Czech Republic\\
}

\email{\authormark{*}kovalenko@optics.upol.cz} 

\begin{abstract}
Two-mode squeezed states are scalable and robust entanglement resources for continuous-variable and hybrid quantum information protocols at a distance. We consider the effect of a linear cross talk in the  multimode distribution of two-mode squeezed states propagating through parallel similar channels. First, to reduce degradation of  the distributed Gaussian entanglement, we show that the initial two-mode squeezing entering the channel should be optimized  already in the presence of a small cross talk. Second, we suggest simultaneous optimization of relative phase between the modes and their linear coupling on a receiver side prior to the use of entanglement, which can fully compensate the cross talk once the channel transmittance is the same for all the modes. For the realistic channels with similar transmittance values for either of the modes, the cross talk can be still largely compensated. This method relying on the mode interference overcomes an alternative method of entanglement localization in one pair of modes using measurement on another pair and feed-forward control. Our theoretical results pave the way to {more} efficient use of multimode continuous-variable photonic entanglement in scalable quantum networks with cross talk.
\end{abstract}

\section{Introduction}
Photonic quantum entanglement \cite{Tittel2001}  is not only a puzzling physical phenomenon, but as well a resource for quantum information processing and quantum communication tasks, in particular for quantum key distribution (QKD) \cite{Gisin2002,Pirandola2019}, quantum metrology \cite{Komar2014} or quantum computing \cite{Kimble2008}.  
Continuous-variable (CV) entanglement, using generally multiphoton states of light, has experimentally enabled quantum communication and information processing with large information capacity \cite{Ukai2011, Su2013, Filip2004, Yokoyama2014, Morin2013, Jeong2014, Yan2017, Huo2018, Huang2019}.  It can be quantified using logarithmic negativity (LN) {\cite{Vidal2002}}, being a measure of {negativity} of a state's partial transpose {\cite{Peres1996,Horodecki1996}}. Advantageously, CV entanglement can be deterministically generated reaching two-mode squeezing bellow  $-10 dB$ (corresponds to logarithmic negativity up to $LN=4.3$)\cite{Eberle2013}. In CV quantum communication, one of the possible ways to simultaneously transfer many entangled states during the same period of time is to use frequency or spatial multiplexing of quantum states residing in different modes  \cite{Hage2010, Heurs2010, Kouadou2019,Shi2020}. Such an approach was in particular used to improve quantum communication with multiplexed QKD \cite{Qu2017} or with multiplexed quantum teleportation using frequency pulse modes \cite{Christ2012}. The techniques for preparation and detection of multimode quantum states of light were drastically improved in the past years, which enabled generation of highly multimode frequency comb states \cite{Valcarcel2006,Pinel2012} with the focus on quantum networks \cite{Roslund2014,Cai2020} or cluster states, which are multiplexed in the time domain  \cite{Yokoyama2013,Yoshikawa2016,Arzani2018,Zhu2020}. 

In a massive mode multiplexing, it is challenging to avoid cross talk between the modes during preparation and distribution \cite{Kudo2014,Szostkiewicz2016,Nada2020}, which can reduce or even destroy logarithmic negativity of entangled pairs and undermine applicability of shared entanglement. It was, in particular, shown that cross talk effects in the multimode homodyne detection can undermine security of mode-non-discriminating CV QKD \cite{Usenko2014,Usenko2015,Kovalenko2019}. However, the cross talk already in the state preparation before the distribution may substantially influence also other entanglement-based protocols, especially, if they are implemented over attenuating (lossy) channels \cite{Pinel2012, Roslund2014}.  

In the current paper, we theoretically study the role of linear cross talk between the signal modes on the state preparation side in distribution of multimode CV entangled states of light through an imperfect (lossy and noisy) channel (differently from the recent study of cross talk from the co-propagating classical signals in multimode CV QKD \cite{Eriksson2019}). We consider Gaussian bipartite two-mode squeezed vacuum states (TMSV) \cite{Weedbrook2012}, also known as twin beams, and show that even a small cross talk substantially degrades Gaussian entanglement contained in the mode pairs and makes the states more sensitive to excess noise in quantum channels used for entanglement distribution. Importantly,  amount of initial two-mode squeezing should be optimally adjusted to maximize its transmission in the presence of cross talk. To overcome this basic passive method, we suggest active control of phase between the modes and controllable linear coupling of the output modes prior to their use on a remote side in order to compensate the cross talk, inspired by the method used to remove correlations in quantum memory channels \cite{Lupo2012} and in the squeezed state generation \cite{Filip2010}. We show that such an active method, if used optimally, can completely eliminate the negative effect of cross talk once the noiseless channel transmittance is the same for all the modes. For a realistic case of asymmetrical quantum channels with slightly  different transmittance for different modes, the cross talk can be still largely compensated for. To demonstrate the advantage we compare our method of active cross talk elimination to a deterministic entanglement localization scheme \cite{Verstraete2004,Sciarrino2009} using homodyne detection on one pair of modes and feed-forward control on another pair. We  show that the active scheme provides better results in a wide range of realistic parameters, particularly, at relatively small cross talk and similar channel transmittance values for different modes, while remarkably preserving multimode structure of the signal. We also address stability of our method and show that it is robust against deviations of the linear coupling used to compensate the cross talk. 

\section{Effect of cross talk on multimode continuous-variable entanglement distribution}

We consider multimode entangled idler and multimode  entangled signal beams constituting TMSV emitted by a source\cite{Weedbrook2012}. Such states are entangled so that quadratures $\hat{x}=\hat{a}^{\dag}+\hat{a}$ and $\hat{p}=i(\hat{a}^{\dag}-\hat{a})$, defined for a given mode, are strongly respectively correlated and anti-correlated within a pair of modes belonging to the signal and the idler beams. To verify the entanglement, a sender (Alice) is performing homodyne quadrature measurement on each of the modes of a respective beam (e.g., signal) and a receiver (Bob) is measuring quadratures of each of many modes of another twin beam (e.g, idler) using another homodyne detector, after the modes experience cross talk and travel through generally attenuating and noisy quantum channels. In the relevant experimental examples \cite{Pinel2012, Roslund2014, Chen2014} the crosstalk appears already in the source prior to the lossy and noisy channel.

\begin{figure}[htb]
\centering\includegraphics[width=0.96\linewidth]{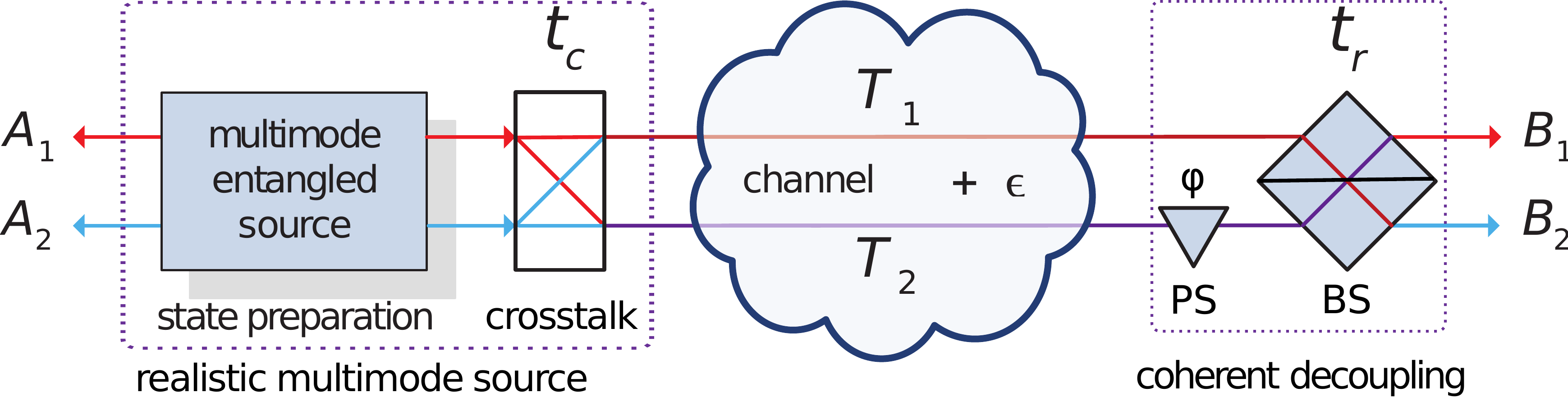}
\caption{Four-mode CV entanglement distribution scheme in the presence of cross talk characterized by the linear coupling $t_c$ between the signal modes, performed over lossy and noisy channels with transmittance $T_i$ for an $i$-th mode and with channel excess noise $\varepsilon$ added to all the modes. In order to  decouple the entangled modes after the channel, we use optical interference, when the remote side optimally applies variable relative phase shift (PS) $\varphi$ between the modes to one of the modes and couples the $B_1$ and $B_2$ modes on a variable beam-splitter (BS) type of interaction with transmittance  $t_r$. We maximize entanglement in pairs of modes $A_1,B_1$ and $A_2,B_2$ over BS transmittance $t_r$ and phase $\varphi$.}
\label{scheme_general}
\end{figure}

We assume the most common linear cross talk between two neighbouring modes, which can be represented by a beamsplitter coupling $t_{c}$ between the modes, as it is schematically shown in Fig. \ref{scheme_general}. {Changing $t_c$ from 0 to 1 then means a transition from very strong inter-mode coupling (hence strong cross talk) to the absence of cross talk. } Here we analyse weak nearest-neighbour coupling, although in general case multimode coupling is more complex, the model allows us to obtain necessary conditions {for improvement to further} numerically apply these methods to a more complex multiple mode cross talk. We parametrize the quantum channel with transmittance $T_i$ for a given $i$-th mode and with the amount of phase-insensitive excess noise $\varepsilon$ added to the quadrature variance, which is assumed (and typically is) the same for all the modes. {Excess noise in general depends on properties of the channel , but can be also concerned with unaccounted detection noise or imperfect state generation. For the fiber channels the observed excess noise is typically below 1 \% shot noise unit (SNU), being the level of vacuum quadrature fluctuations, at the channel output  \cite{Jouguet2012, Huang2016}.}%
We first analyse the role of cross talk in a basic case of two entangled mode pairs (so that the overall number of modes in signal and idler beams is four), then extend the results to three pairs. 

We describe the Gaussian multimode TMSV states \cite{Weedbrook2012} by covariance matrices with elements $\gamma_{ij}=\langle r_i r_j \rangle$, {where} $r_i=\{x_i,p_j\}$ is a set of quadratures of a given $i$-th mode, taking into account their zero mean values,  $i \in [1,2N]$, here $2N$ is the overall number of modes (hence, $N$ is the number of mode pairs, or, equivalently, the number of modes in each signal and idler beams). Then the system of four pairwise {perfectly} entangled modes $A_{\{1,2\}},B_{\{1,2\}}$, shown in Fig. \ref{scheme_general}, each having quadrature variance $V \geq 1$,  is described by the 8x8 covariance matrix of the form

\begin{equation}\gamma_{A_1A_2B_1B_2}=
\left(
\begin{smallmatrix}
 V~\mathbb{I} &  \sqrt{V^2-1} ~\mathbb{Z} & 0 & 0  \\
  \sqrt{V^2-1}  ~\mathbb{Z} & V~\mathbb{I} & 0 & 0\\
 0 & 0  & V~\mathbb{I} &  \sqrt{V^2-1}  ~\mathbb{Z}\\
 0  & 0  & \sqrt{V^2-1} ~\mathbb{Z} & V  ~\mathbb{I}\\
\end{smallmatrix}
\right).
\label{covmat}
\end{equation}
Here $\mathbb{I} =  { diag[1,1]}$ is the unity matrix, $\mathbb{Z}= {diag[1,-1]}$ is a Pauli matrix. Quadrature variance $V$ is related to the squeezing parameter $r$ of TMSV, which defines the amount of two-mode squeezing applied to the two-mode vacuum in both quadratures  { $\text{Var}[(x_{A_i}+x_{B_i})/\sqrt{2}]=e^{-2r}$ and $\text{Var}[(p_{A_i}-p_{B_i})/\sqrt{2}]=e^{-2r}$}, and $V=\cosh(2r)$, so that the larger $r$, the stronger is quadrature correlations between the modes, given by $\langle (x_{A_i}+x_{B_i}) (p_{A_i}-p_{B_i}) \rangle=\sinh{r}$.
After a linear cross talk $t_c$ and lossy and noisy quantum channel,  the covariance matrix changes to
\begin{equation}\gamma_{A_1A_2B_1B_2}=
\left(
\begin{smallmatrix}
 V~\mathbb{I} & \sqrt{t_c ~T_1} \sqrt{V^2-1} ~\mathbb{Z} & 0 & -\sqrt{r_c~T_2} \sqrt{V^2-1}  ~\mathbb{Z} \\
 \sqrt{t_c~T_1} \sqrt{V^2-1}  ~\mathbb{Z} & [T_1 (V+\varepsilon-1)+1]~\mathbb{I} & \sqrt{r_c~T_2} \sqrt{V^2-1} ~\mathbb{Z} & 0\\
 0 & \sqrt{r_c~T_1} \sqrt{V^2-1} ~\mathbb{Z} & V~\mathbb{I} & \sqrt{t_c~T_2} \sqrt{V^2-1}  ~\mathbb{Z}\\
 -\sqrt{r_c~T_1} \sqrt{V^2-1} ~\mathbb{Z} & 0  & \sqrt{t_c~T_2} \sqrt{V^2-1} ~\mathbb{Z} & [T_2 (V+\varepsilon-1)+1]  ~\mathbb{I}\\
\end{smallmatrix}
\right)
\label{covmatct}
\end{equation}
Here $r_c \equiv 1-t_c$.
{We characterize the bipartite Gaussian entanglement using LN \cite{Vidal2002} of a state $\rho$ is defined as
\begin{equation}
 LN(\rho) =- \log_2 ||\rho^\Gamma||_1 , 
 \label{eq:lnrho}
\end{equation}
where  $\rho^\Gamma$ is a partial transpose of $\rho$, $||\rho||_1$ is the trace norm of the operator $\rho$, that is equal to the sum of the absolute values of the negative eigenvalues of $\rho^\Gamma$, quantifying the degree to which a partially transposed state fails to be
positive \cite{Peres1996,Horodecki1996}. For a  Gaussian states with covariance matrix $\gamma$ the eq. (\ref{eq:lnrho}) becomes the sum of all symplectic eigenvalues of the partially transposed that are less than one: $LN(\gamma) = \sum_k \max \lbrace 0, - \log_2 \nu_k\rbrace $. For the TMSV states with the symplectic eigenvalues $\{\nu_+, \nu_-\}$ the larger eigenvalue is $\nu_+\geqslant 1$ \cite{Adesso2004}. This way the LN of the $i$-th pair of modes simplifies to }
\begin{equation}
 LN_i =\max \lbrace 0, - \log_2 \nu_{i_-} \rbrace, 
\end{equation}
where $\nu_-$ is the smallest symplectic eigenvalue of the covariance matrix of the partially transposed state of either the first or the second pair. {We limit our study to the Gaussian TMSV states as the typical case of bipartite quadrature-entangled states of light.}  As the task is to deterministically transmit all pairs through the channel, we evaluate logarithmic negativity for each pair of modes separately. We also define the {\it initial logarithmic negativity} of a TMSV state (\ref{covmat}), before the cross talk and before sharing the state over a noisy and lossy channel, see Fig.\ref{scheme_general}, as
\begin{equation}
LN_0(V) =-\frac{1}{2 }\log_2 \Big(2 V^2-1-2 V \sqrt{V^2-1}\Big),
\end{equation}
which is the same for the both pairs of modes. It represents the maximum of Gaussian entanglement in the pairs of modes, which can be shared in the perfect case of no cross talk and ideal (lossless and noiseless) channels. In the limit of small state variance ($V\to 1$), it behaves as $LN_0\sim \frac{1}{\log 2} \sqrt{2(V-1)} $. In the limit of large $V \to \infty$, $LN_0 \sim\log_2 (2V)$.

In the absence of cross talk ($t_c = 1$) the logarithmic negativity monotonously grows with increasing TMSV quadrature variance $V$, but its limit depends on the channel parameters as
\begin{equation}
\lim_{V\to\infty} LN  = -\log_2 \frac{1-T_i(1-\varepsilon)}{1+T_i}
\label{eq:lnInfty}
\end{equation}
For noiseless channels ($\varepsilon =0$) the entaglement never vanishes for any $T_i>0$. For lossless channels ($T_i=1$) entanglement {is lost at} $\varepsilon > 2$. 
In the presence of cross talk between the signal modes and after {a lossy and noisy} channel, as shown in Fig. \ref{scheme_general}, the logarithmic negativity becomes 
\begin{equation}
\begin{split}
\label{eq:lni}
LN =-\frac{1}{2} \log _2\frac{1}{2} \Big(1+2 T_i[\varepsilon+(V-1) (t_c V+t_c+1)]+T_i^2 (\varepsilon+V-1)^2+V^2-\\
[1+V+T_i(\varepsilon +V-1)] \sqrt{T_i^2 (\varepsilon +V-1)^2+(V-1)^2 -2 T_i(V-1) [\varepsilon -2 t_c (V+1)+V-1]}\Big)
\end{split}
\end{equation}
For channels with high loss {($T_i\ll 1$)},  expanding (\ref{eq:lni}) around $T_i=0$ gives
\begin{equation}
\label{eq:lniT=0}
LN \approx \frac{T_i [2-\varepsilon-(1-t_c) (V+1)] }{\log (2)}\left(1+\frac{T_i}{2}[2-\varepsilon-(1-t_c) (V+1)]-\frac{T_i t_c (V+1)}{V-1}\right)
\end{equation}
For a very small transmittance{, $T_i\to 0$, LN is well described by the first term in} (\ref{eq:lniT=0}), $LN \sim \frac{T_i [2-\varepsilon-(1-t_c) (V+1)] }{\log (2)}$. In this limit, the shared entanglement has linear dependence on the transmittance $T_i$. Provided that the cross talk is absent ($t_c=1$), {the entanglement is} independent on the initial variance $V$ and is destroyed by the excess noise $\varepsilon=2$. The presence of cross talk introduces a new term $ -T_i(1-t_c) V$ that reduces the shared entanglement, which then monotonically {decreases} with the {increase} of $V$.

\begin{figure}[htbp]
\begin{minipage}{0.5 \linewidth}
\includegraphics[width=0.95\linewidth]{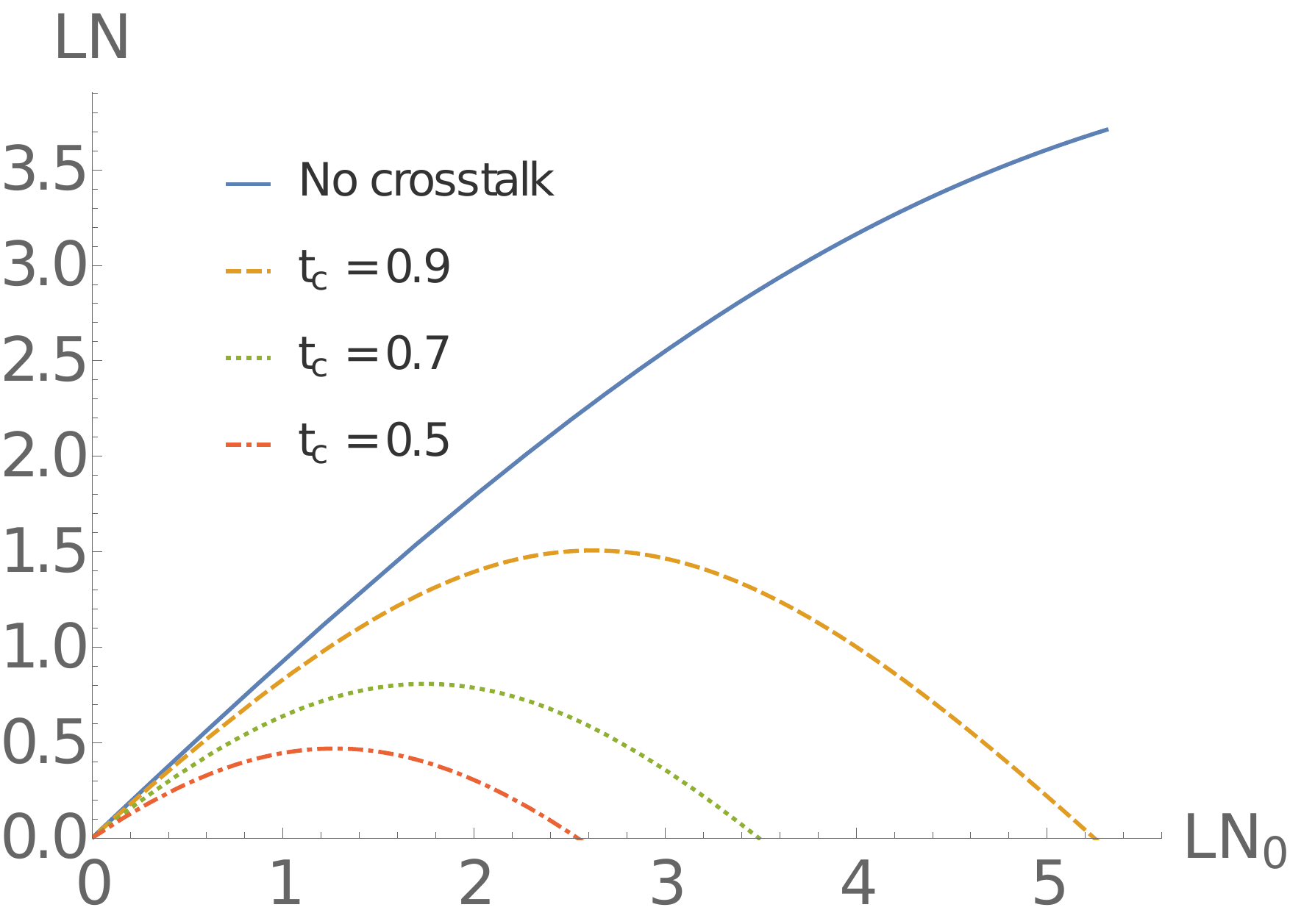}
\end{minipage}
\hfill
\begin{minipage}{0.5 \linewidth}
\includegraphics[width=0.95\linewidth]{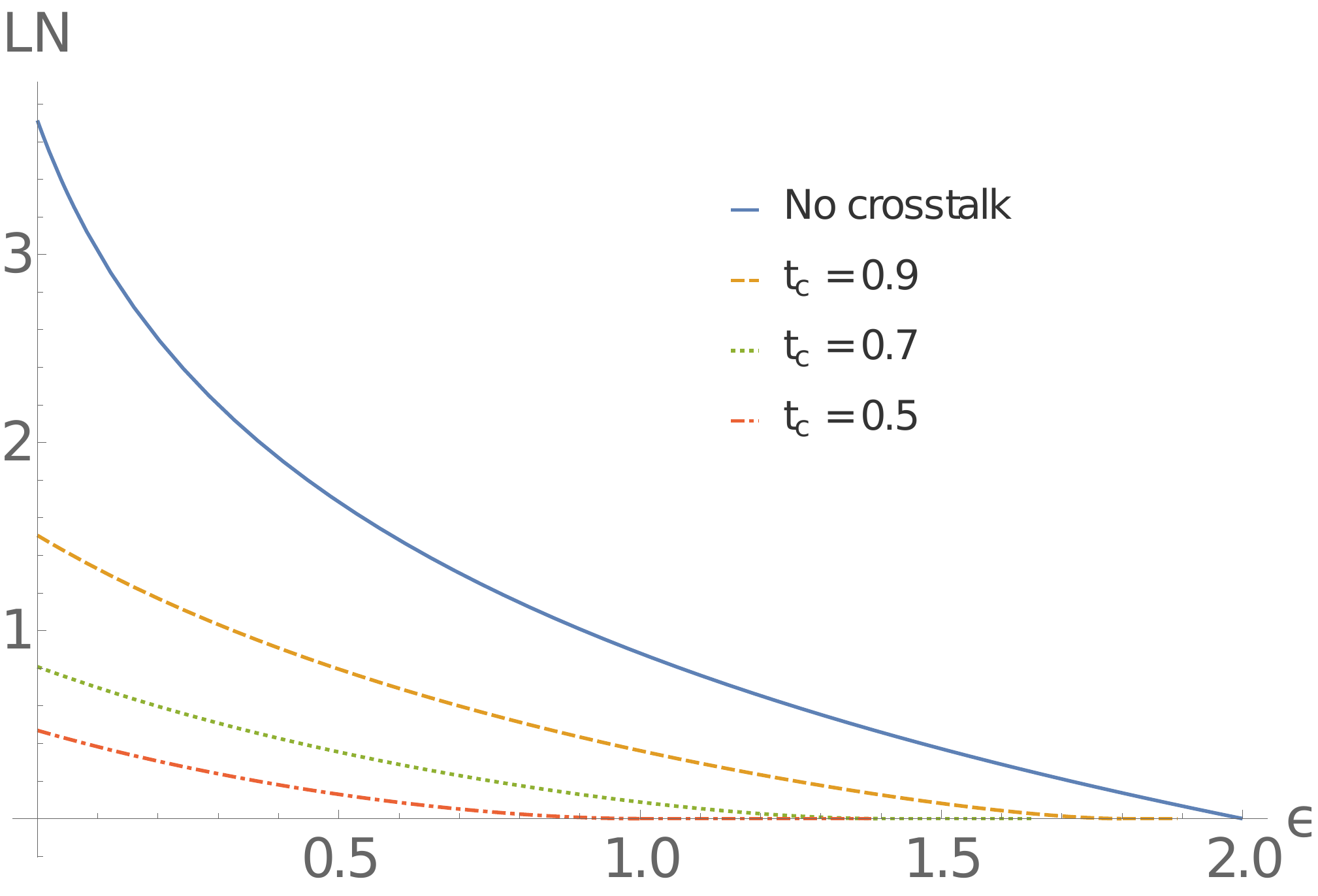}
\end{minipage}
\caption{Gaussian logarithmic negativity of the pair of modes $A_1,B_1$ after cross talk $t_c$ and after passing through a quantum channel versus the initial logarithmic negativity of the state (left) and versus the channel noise (right). The channel transmittance for both the modes is $T_{1,2}=0.9$. Left: no excess noise ($\varepsilon=0$). Right: fixed state variance $V=5$. Entanglement is evidently reduced and made more sensitive to the excess noise even by a small cross talk, and is destroyed completely by the excess noise reaching {the} threshold (\ref{eq:emax}). The initial entanglement can be optimized as in (\ref{eq:vopt}) to reach the maximum shared entanglement.}
\label{ln(v)} 
\end{figure}

Beyond the limit of small initial entanglement ($V \rightarrow 1$), sensitivity of Gaussian entanglement of TMSV to  cross talk can be seen in Fig. \ref{ln(v)}, left, where logarithmic negativity is plotted versus the initial entanglement at different cross talk coupling $t_c$ values and compared to the case when {the} cross talk is absent ($t_c=1$).
It is numerically evident from (\ref{eq:lni}) and from the plots in Fig. \ref{ln(v)}, left, that the amount of entanglement shared over the noiseless channel in the presence of cross talk becomes sensitive to the initial entanglement even for weak channel attenuation $T_i=0.9$ and can be broken once the initial entanglement is too strong. This is concerned with the fact that due to cross talk the signal from an adjacent mode is coupled to the signal in another mode and thus effectively acts an uncorrelated noise appearing in one of the beams if all modes are treated independently. Amount of such a noise is larger for the higher state variance, i.e., for a larger initial entanglement. This destructive extra noise effect from the crosstalk is also dominant for channels with excess noise. Moreover, the presence of cross talk degrades robustness of the shared entanglement to the channel noise, as can be seen from Fig. \ref{ln(v)}, right.  
 The shared entanglement is then destroyed by the level of noise, which is independent of the channel transmittance $T_i$ (hence being valid for any $T_i>0$), and is  given by 
 \begin{equation}
\label{eq:emax}
\varepsilon_{max} = 1+t_c-(1-t_c)V.
\end{equation}
Equivalently, the bound on the shared entanglement (variance $V$ for which entanglement turns to $0$) is the same for any $T_i>0$, and it reads
\begin{equation}
\label{eq:vmax}
 V_{max} = \frac{1+t_c- \varepsilon}{1-t_c}
%
\end{equation}
 Importantly, the maximum tolerable initial variance given by $V_{max}$ (or, equivalently, maximum tolerable initial entanglement) {also} does not depend on the channel transmittance and, for a purely attenuating channel, depends only on the cross talk coupling $t_c$. The necessary condition $V<V_{max}$  therefore simplifies optimization if the cross talk is present only in the source.  Note that if the channel is present also before the crosstalk, optimization becomes more complex and involves the channel transmittance values, however, the limitation $V<V_{max}$ remains. 

The above given results were obtained in the assumption that {the} cross talk occurs between two modes. In the case, when each of the multiple TMSV modes interacts with two neighbouring modes, i.e., the cross talk dominantly occurs between three modes, the equations (\ref{eq:lni}, \ref{eq:vmax}) hold up to the substitution $t_c \to t_c^2$, which leads to stronger sensitivity of the shared entanglement to the linear cross talk between the signal modes. In this case, $V_{max}$ is even more limited and therefore less entanglement can be transmitted. 

The shared entanglement, in the presence of cross talk, can be maximized by optimizing the initial entangled resource. The optimal state variance $V$, which maximizes the shared entanglement, reads
\begin{equation}
\label{eq:vopt}
V_{opt} = \frac{ (1-t_c) (1-T_i) (1-T_i+\varepsilon T_i) + (T_i+1) \sqrt{(1-t_c) t_c T_i  [4 t_c -\varepsilon (2-2T_i+\varepsilon T_i)]}}{(1-t_c) [1+T_i (4 t_c+T_i-2)]}
\end{equation}
By optimizing the initial TMSV variance as in (\ref{eq:vopt}) we both reduce the negative influence of the cross talk, and rise tolerance to the excess noise. The maximal tolerable level of excess noise then becomes $\varepsilon_{max}=2 t_c $ SNU.

For a small cross talk $t_c \to 1$ the optimal variance can be approximated as
\begin{equation}
\label{eq:voptt=1}
V_{opt} \approx \frac{(1-T_i) (1-T_i+\varepsilon T_i)}{(1+T_i)^2}+\frac{\sqrt{(2-\varepsilon)T_i (2+\varepsilon T_i)}}{(1+T_i)\sqrt{1-t_c}}.
\end{equation}
If we neglect the excess noise, the expression (\ref{eq:voptt=1}) simplifies to
\begin{equation}
\label{eq:voptt=1e=0}
V_{opt} \approx \left( \frac{1-T_i }{1+T_i}\right) ^2+\frac{2 \sqrt{ T_i}}{(1+T_i)\sqrt{1-t_c}}.
\end{equation}
Putting $T_i=1$ in (\ref{eq:voptt=1e=0}) allows to easily demonstrate the existence of the upper bound on  the optimal state variance $V_{opt}(T=1)=\frac{1}{\sqrt{1-t_c}}$. It illustrates the fact that for any cross talk, even without noise and attenuation, the maximal possible shared entanglement is rather limited.
For a strongly attenuating channel ($T_i \to 0$) together with excess noise $\varepsilon$, the optimal variance $V_{opt}$ is approximately
\begin{equation}
V_{opt} \approx 1+\frac{\sqrt{2 T_i t_c (2 t_c-\varepsilon)}}{\sqrt{1-t_c}}-(4 t_c-\varepsilon)T_i
\label{eq:voptT=0}
\end{equation}
which means that the optimal initial quadrature variance of TMSV is rather small (close to one) when channel is strongly attenuating even if the cross talk is small ($t_c \to 1$).
Therefore, amount of entanglement, which can be shared over lossy channels in the presence of even small cross talk, is then strongly limited.
Note, that (\ref{eq:voptT=0}) explicitly shows the condition on excess noise $\varepsilon$ for the optimized $V$  being  $\varepsilon < 2 t_c~SNU $.

For a channel with strong attenuation ($T_i \to 0$), the optimized entanglement can be {then} approximated as
\begin{equation}
LN (V_{opt}) \approx \frac{2 T_i\left[t_c-\frac{\varepsilon}{2}- 2 \sqrt{ t_c (1-t_c)(t_c-\frac{\varepsilon}{2})T_i}+[3 t_c(1-t_c)+\frac{\varepsilon}{2}(1-\frac{\varepsilon}{2})] T_i \right]}{\log (2)}
\label{eq:lnT=0}
\end{equation}
or, in the absence of excess noise, 
\begin{equation}
LN (V_{opt}) \approx \frac{2 t_c T_i \left(1-2 \sqrt{(1-t_c)T_i} +3 (1-t_c) T_i\right)}{\log (2)}.
\label{eq:lnT=0e=0}
\end{equation}

In the limit of a very small cross talk ($t_c >0.98$) for relatively strong excess noise ($\varepsilon>0.2$ SNU) it is possible to find better approximation expanding optimized entanglement $LN(V_{opt})$ into series for  a very low cross talk ($t_c \sim 1$) in the the limit of strongly attenuating channel ($T_i \to 0$), then the optimized entanglement can be better analytically approximated as
\begin{equation}
LN (V_{opt})  \approx -\log _2\left(\frac{1-T_i+\varepsilon T_i}{1+T_i}\right) -\frac{2 T_i \sqrt{(2-\varepsilon) T_i (2+\varepsilon T_i)}}{(1+T_i) (1-T_i+\varepsilon T_i)\log (2) }\sqrt{1-t_c}
\label{eq:lnt=1T=0}
\end{equation}

\begin{figure}[htbp]
\begin{minipage}{0.5 \linewidth}
\includegraphics[width=0.95\linewidth]{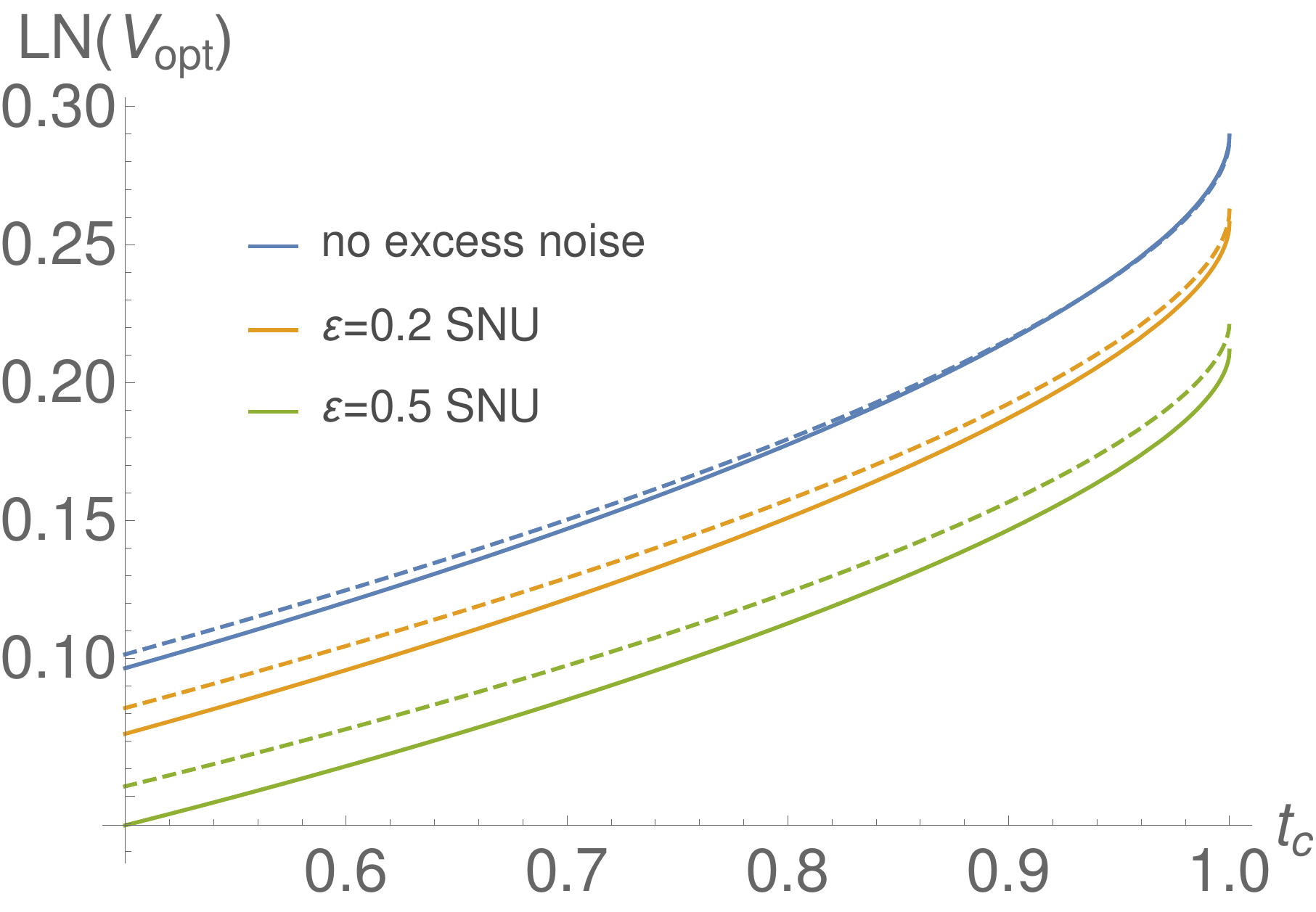}
\end{minipage}
\hfill
\begin{minipage}{0.5 \linewidth}
\includegraphics[width=0.95\linewidth]{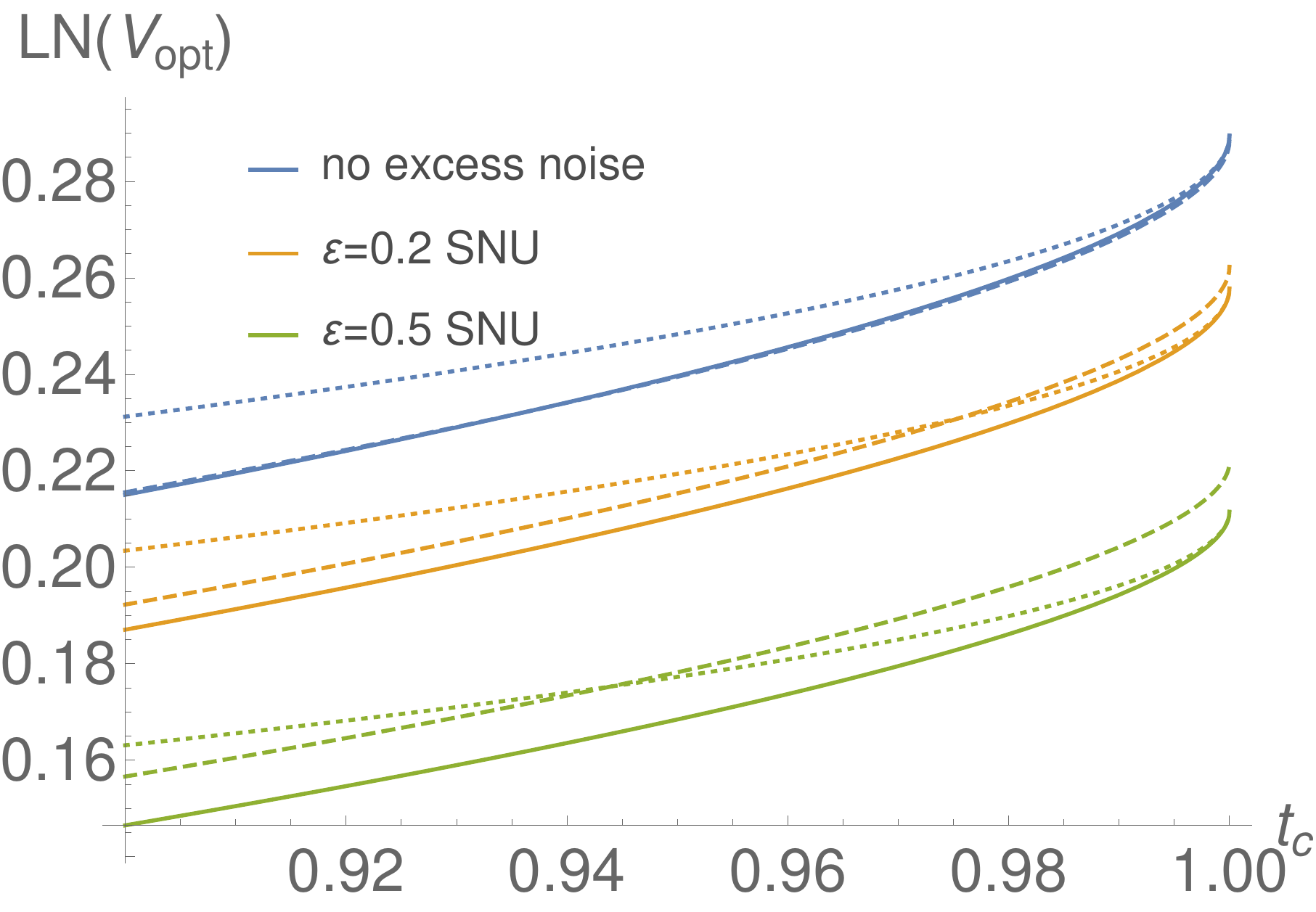}
\end{minipage}
\caption{Optimized logarithmic negativity in one pair of modes versus cross-talk coupling $t_c$ after passing through the attenuating channel of transmittance $T_i=0.1$ for different levels of excess noise. The initial variance of the state is optimized. Solid lines correspond to the exact results, dashed lines -- to the approximation (\ref{eq:lnT=0}), dotted lines -- to the approximation (\ref{eq:lnt=1T=0}). Entanglement is reduced by cross-talk quite strongly even when the state is optimized before sharing.}
\label{lnopt(tc)} 
\end{figure}
Figs. \ref{ln(v)} -- \ref{lnopt(tc)}  illustrate that although optimization of the state variance during the state preparation helps to somewhat mitigate the negative effects of cross talk (allowing to reach maximal logarithmic negativity for given cross talk and excess noise that corresponds to the maximums in Fig. \ref{ln(v)}), nevertheless such optimization fails to fully compensate for the cross cross talk especially in the presence of a stronger excess noise. Comparison of the left and the right panel in the Fig. \ref{lnopt(tc)} and the eqs. (\ref{eq:lnT=0}-\ref{eq:lnt=1T=0}) show that in the strongly attenuating channel while for a weak cross talk logarithmic negativity decreases as $LN\sim -\sqrt{1-t_c}$ (eq. (\ref{eq:lnT=0}) and  {dashed} lines), very soon it gets suppressed more by higher orders of $t_c$ (eq. (\ref{eq:lnT=0}-\ref{eq:lnt=1T=0}) and  {dotted} lines).
Thus, even though initial entanglement can be optimized to achieve the maximum shared entanglement in the presence of cross talk, the shared entanglement will be still largely  reduced by the cross talk, especially if combined with {the} channel loss and/or noise.

We therefore consider the local active manipulations on either sender or receiver side in order to improve and possibly fully restore the shared entanglement by compensating the cross talk.  If the mode coupling happens only in the source during the state preparation, it can be compensated by manipulating either the modes $A_1$ and $A_2$, or $B_1$ and $B_2$  on the sender side or the modes $B_1$ and $B_2$ on the remote side after the states are shared through an imperfect channel. If an additional mode coupling also happens in the channel, manipulations on the receiver side are needed in addition to the manipulations at the sender side in order to simultaneously compensate both cross talk in the state preparation and in the channel. Generally both, the cross talk and the losses, can occur on both sender and receiver side and also in the source itself during the state preparation and distribution. In our study we assume that the shared part of the states (hence, modes $B_1$ and $B_2$) is the subject to cross talk. We then consider the manipulations on the receiver side, i.e., after the imperfect channel, in order to compensate the cross talk. The case of an ideal cross talk compensation on the sender side would be equivalent to the absence of cross talk, as can be seen in the next section.

\section{Cross talk compensation by optical interference}

We consider the feasible state operations prior to using the entangled states in order to compensate the cross talk, i.e., to reduce or completely eliminate its negative effect on the entanglement distribution. Since the cross talk is typically caused by  an energetically passive photon exchange between the signal modes, we consider the beam-splitter type interaction based on a variable coupling $t_r$ between the signal modes.  The compensating interaction can in principle be performed on either sender or receiver side. If  no signal loss occurs on the sender side, it is sufficient to implement the beam-splitter with  $t_r=t_c$ coupling the modes $A_1$ and $A_2$ at the sender. To take into account the effects caused by the lossy channels we further concentrate on the case where the compensation is implemented on the modes $B_1$ and $B_2$ by the remote party of the entanglement distribution scheme.  In this case the beam splitter needs to be preceded by  an optimal phase  shift (in  the case of linear cross talk between two modes  it is given by $\pi$) on one of the modes (e.g., mode 2), as shown in Fig. \ref{scheme_general}. Then for a pair $A_1,B_1$ the covariance matrix reads
\begin{small}
\begin{equation}
\gamma_{A_1B_1}=
\left(
\begin{array}{cc}
 V ~\mathbb{I} & \left(\sqrt{T_2 r_c r_r}+\sqrt{T_1 t_c t_r}\right) \sqrt{V^2-1}  ~\mathbb{Z} \\

 \left(\sqrt{T_2 r_c r_r}+\sqrt{T_1 t_c t_r}\right) \sqrt{V^2-1}  ~\mathbb{Z} & [1+T_1 t_r (V-1)+T_2 r_r(V-1)] ~\mathbb{I}\\

\end{array}
\right),
\label{covmatrev}
\end{equation}
\end{small}
where $r_c\equiv 1-t_c$, $r_r\equiv 1-t_r$, and similarly for $A_2,B_2$ up to the replacement $T_1 \leftrightarrow T_2$.
It follows from the expression (\ref{covmatrev}), that for the perfectly balanced  noiseless channels $T_1=T_2 \equiv T$ putting $t_r=t_c$ turns the correlations into $\left(\sqrt{T_2 r_c r_r}+\sqrt{T_1 t_c t_r}\right)\sqrt{V^2-1}=\left(\sqrt{T}(1- t_c)+\sqrt{T}t_c\right)\sqrt{V^2-1}=\sqrt{T}\sqrt{V^2-1}$ and the variances into $1+T t_r (V-1)+T(1- t_r)(V-1)= 1+T (V-1)$ so that the covariance matrix  turns into the one without any cross talk:
\begin{small}
\begin{equation}
\gamma_{A_1B_1}=
\left(
\begin{array}{cc}
 V ~\mathbb{I} & \sqrt{T } \sqrt{V^2-1}  ~\mathbb{Z} \\
\sqrt{T } \sqrt{V^2-1}  ~\mathbb{Z} & [1+T (V-1)] ~\mathbb{I}\\

\end{array}
\right),
\label{covmatrevperf}
\end{equation}
\end{small}
similarly for $A_2,B_2$, which fully restores entanglement and other Gaussian characteristics, affected by the cross talk. This can be generalized to {a pair of} arbitrary pure two-mode states in modes {$A_1, A_2$ and} $B_1,B_2$, considered {as the product of $\iint d^2 \bar{\alpha} d^2 \alpha~ P(\bar{\alpha},\alpha) |\bar{\alpha}\rangle_{A1} |\alpha\rangle_{B1}$ 
and 
$\iint d^2 \bar{\beta} d^2 \beta ~P(\bar{\beta},\beta) |\bar{\beta}\rangle_{A2} |\beta\rangle_{B2}$} in the coherent-state overcomplete basis. { This overcomplete basis decomposing unity allows to straightforwardly calculate the result for any pure two-mode state in equally lossy channels.} Indeed, if two coherent {basis} states {$|\alpha\rangle_{B_1}$ and $|\beta\rangle_{B_2}$} experience a linear cross talk $t_c$ and pass through the attenuating channels $T_1$ and $T_2$ and through the decoupling scheme with a phase shift and coupling $t_c$ as shown in Fig. \ref{scheme_general}, disregarding normalization and general phase they are transformed as
{
\begin{equation}
|\alpha'\rangle_{B_1} = \left| \left(\sqrt{T_1 t_c t_r}+\sqrt{T_2 r_c r_r}\right)\alpha+\left(\sqrt{T_2 t_c r_r}-\sqrt{T_1 r_c t_r}\right)\beta \right\rangle_{B_1}
\end{equation}
and
\begin{equation}
|\beta'\rangle_{B_2} = \left|\left(\sqrt{T_1 r_c r_r}+\sqrt{T_2 t_c t_r}\right)\beta +\left(\sqrt{T_2 r_c t_r}-\sqrt{T_1 t_c r_r}\right)\alpha \right\rangle_{B_2}.
\end{equation}
}
Choosing $t_r=\frac{T_2 t_c}{T_2 t_c+ T_1 r_c}$  the states change to
{
\begin{equation}
|\alpha''\rangle_{B_1} = \left| \frac{\sqrt{T_2 T_1  }}{\sqrt{T_2 t_c+ T_1 r_c}} \alpha \right\rangle_{B_1}
\label{eq:alphaFin}
\end{equation}
\begin{equation}
|\beta''\rangle _{B_2}= \left|\sqrt{T_2 t_c+ T_1 r_c}\beta +\frac{(T_2-T_1)\sqrt{t_c r_c}}{\sqrt{T_2 t_c+ T_1 r_c}} \alpha \right\rangle_{B_2}.
\label{eq:betaFin}
\end{equation}}
We have therefore eliminated the contribution of the second mode {$B_2$} to the first one {$B_1$}, but not  vice versa. If, on the other hand,  we chose $t_r=\frac{T_2 t_c}{T_2 t_c+T_1 r_c} $, the cross talk will be eliminated from the second mode, but not from the first one.
It is easy to see, that if $T_1=T_2 \equiv T $, (\ref{eq:alphaFin}) and (\ref{eq:betaFin}) turn into {$|\alpha''\rangle_{B_1} = \left| \sqrt{T}\alpha \right\rangle_{B_1} $ and $|\beta''\rangle _{B_2}= \left|\sqrt{T }\beta \right\rangle_{B_2}$ }as it should be for two independent modes attenuated by a channel. Therefore for  the perfectly balanced channels ($T_1=T_2$) the cross talk is fully eliminated for {a pair of} arbitrary pure two-mode state.

\begin{figure}[htbp]
\begin{minipage}{0.49 \linewidth}
\includegraphics[width=0.99\linewidth]{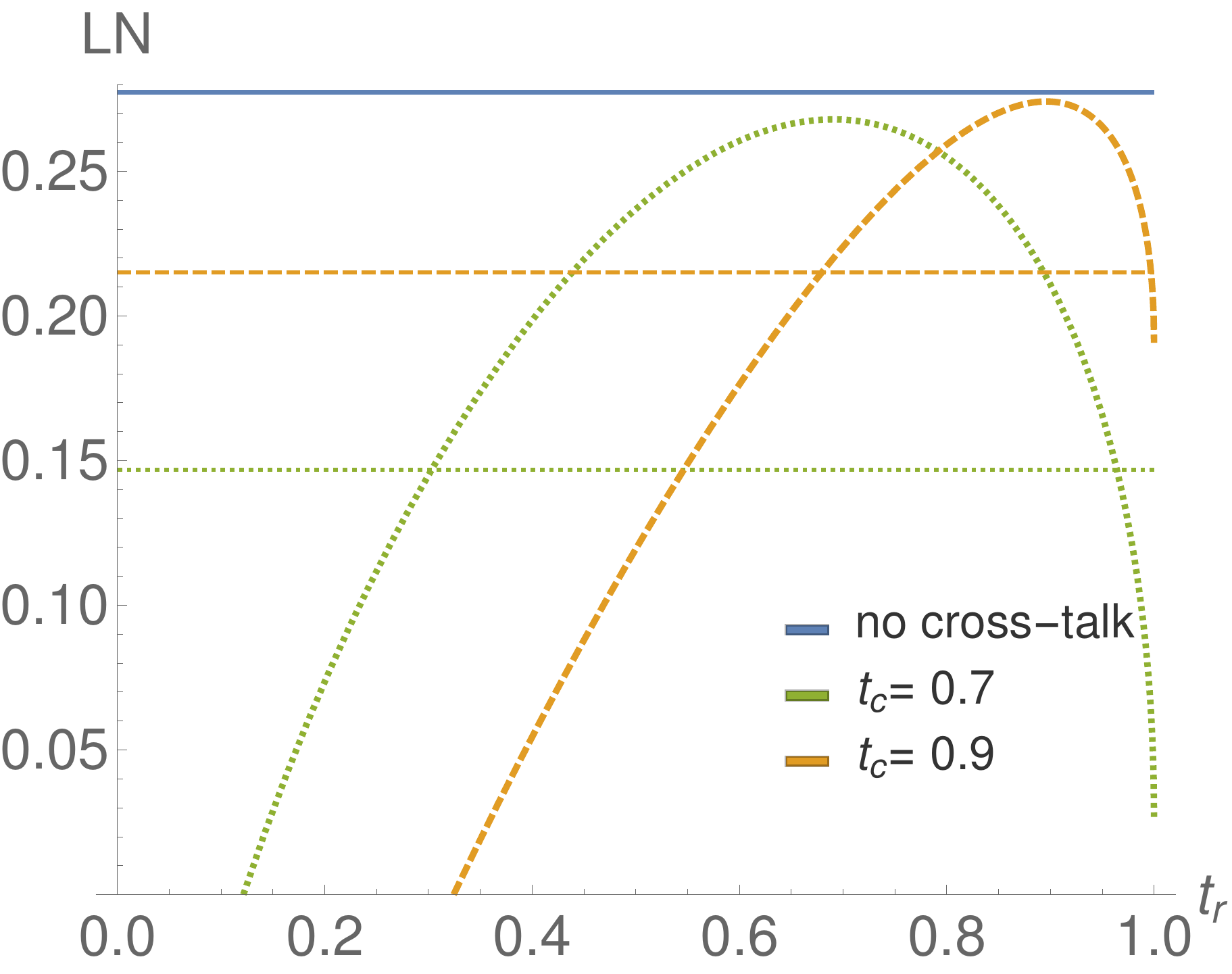}
\end{minipage}
\hfill
\begin{minipage}{0.49 \linewidth}
\includegraphics[width=0.99\linewidth]{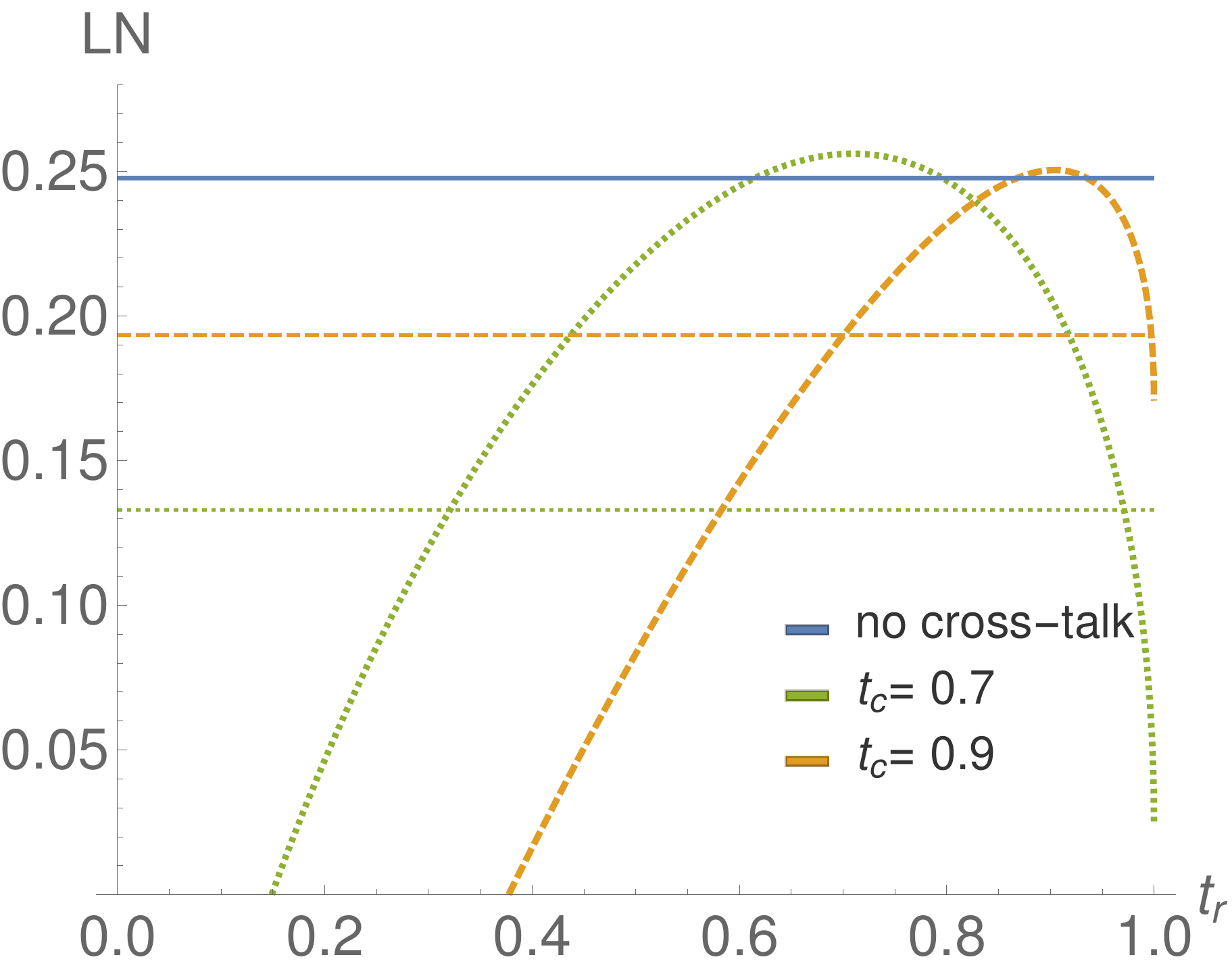}
\end{minipage}
\caption{Logarithmic negativity  versus reverse coupling $t_r$ aimed at compensating the cross talk after the unbalanced channels with the imbalance $T_1=-10 dB$ and $T_2=-10.5 dB$  of the first pair of modes ($A_1B_1$, left), the second pair ($A_2B_2$, right). The uppermost (blue) horizontal line shows the logarithmic negativity of the respective pair without the cross talk, the upper (orange) and the lower (green) curves show the result of applying compensating beam-splitter of transmittance $t_r$ on Bobs's side at the same cross talk parameter $t_c$. Initial state variance is $V = 5$, no excess noise ($\varepsilon=0$).  The dotted horizontal lines (orange and green) correspond to the logarithmic negativity without the cross talk compensation but with the initial entanglement optimization.}
\label{reversing}
\end{figure}

When the channel transmittance values $T_1,T_2$ for modes $B_1,B_2$ are different, the cross talk cannot be fully compensated  in both modes and entanglement cannot be fully restored  in both transmitting channels. {While transmittance values for different modes are typically similar in fiber channels \cite{Alwayn2014}, they can vary, e.g. for frequency modes in atmospheric channels \cite{Streete1968}. We therefore consider the efficiency of the cross talk compensation in the unbalanced channels.} In order to maximally reconstruct entanglement, one has to optimize the coupling $t_r$, as shown in Fig. \ref{reversing}.
It is evident from the plots, that entanglement can still be largely compensated for  weaker crosstalk  ($t_c\to 1$) between asymmetrical lossy channels and that the method of optical interference is stable to the setting $t_r$.  There is a difference between the optimal $t_r$ for the first  pair and for the second  one  (it corresponds to the maximums of the curves on  {the} left and right plots in Fig. \ref{reversing}), although, as it noticeable from the plots, their maximums are very close.  In the second  pair  $A_2B_2$ we even can overcome the entanglement before the  cross talk because we get stronger signal in the  pair $A_2B_2$ that is taken from the  pair $A_1B_1$  due to the cross talk. 

In a general case, $t_r$ should be optimized numerically to result in the maximally restored entanglement. However, the optimal setting can be derived analytically in the relevant limits of small and large initial entanglement (or, equivalently, small and high large initial state variance $V$). For a purely lossy channel ($\varepsilon=0$),  the optimal transmittance, that maximizes the entanglement in the first pair of modes $A_1B_1$, reads
\begin{equation}
t_r = \frac{T_1 t_c}{T_1 t_c+ T_2 (1-t_c)}, ~~V \approx 1
\label{tr1}
\end{equation}
and
\begin{equation}
t_r = \frac{T_2 t_c}{T_2 t_c+T_1 (1-t_c)}, ~~V\to\infty
\label{trInf}
\end{equation}
 And vice versa ($T_1\leftrightarrow T_2$) for another pair. It is important to note that the optimal $t_r$ for any value of the state variance $V$ always lies between the bounds given by (\ref{tr1}-\ref{trInf}) and, for realistically close $T_1$ and $T_2$, this  {interval} is quite narrow. 
For $t_r$ defined by (\ref{trInf}) the logarithmic negativity is an increasing function of $V$ and for large initial state variance it approaches the limit:
\begin{equation}
\lim_{V\to\infty} LN_{t_r} = -\log_2 \left[\frac{t_c T_2+T_1 (1-t_c-T_2)}{t_c T_2+T_1 (1-t_c+T_2)}\right].
\label{eq:ln_decoupl(inf)}
\end{equation}
Note, that the above given equations (\ref{tr1}-\ref{eq:ln_decoupl(inf)})  {are applicable} for maximization of the enanglement for the pair $A_1,B_1$. If, on the other hand, the goal is to maximize the logarithmic negativity in the second pair  $A_2,B_2$ , the same equations (\ref{tr1}-\ref{eq:ln_decoupl(inf)}) can be applied with replacement $T_1 \leftrightarrow T_2$. This also implies, that in the case of unbalanced channels $T_1 \neq T_2$,  {the} entanglement cannot be optimally restored  in both modes. While applying  {the} optical interference method to compensate  the cross talk in  the channels with different transmittance for each mode ($T_1 \neq T_2$), one has to chose  one mode in which to maximize the entanglement at the expense of the rest of the modes. The equations (\ref{tr1}-\ref{eq:ln_decoupl(inf)}) refer to the  first pair of modes $A_1B_1$.   However, since in practical situations strong unbalancing between transmittance values for different modes is unlikely, i.e. $T_1$ is typically close to $T_2$, optimal settings for $t_r$ are also close and choosing $t_r$ in between the optimal settings for either of the pairs will give nearly optimal results in terms of entanglement for both the pairs.

Advantage of the decoupling by interference is that cross talk from {\em all} entangled pairs can be nearly perfectly removed.  Moreover, technical imperfections, as the modes match imperfectly at the decoupling BS, can be incorporated to the full optimization.  Despite such a  principal advantage, one may be interested in  maximizing entanglement in only one pair of modes. In this case entanglement  localization  \cite{Verstraete2004,Sciarrino2009} can be applied by performing only an optimized measurement on one pair and feed-forward control on another pair of modes based on the measurement outcomes, which we consider in the next Section and compare it to the above suggested method of reversed coupling.

\section{Entanglement  localization by measurement and feed-forward control}

\begin{figure}[htb]
\centering\includegraphics[width=0.96\linewidth]{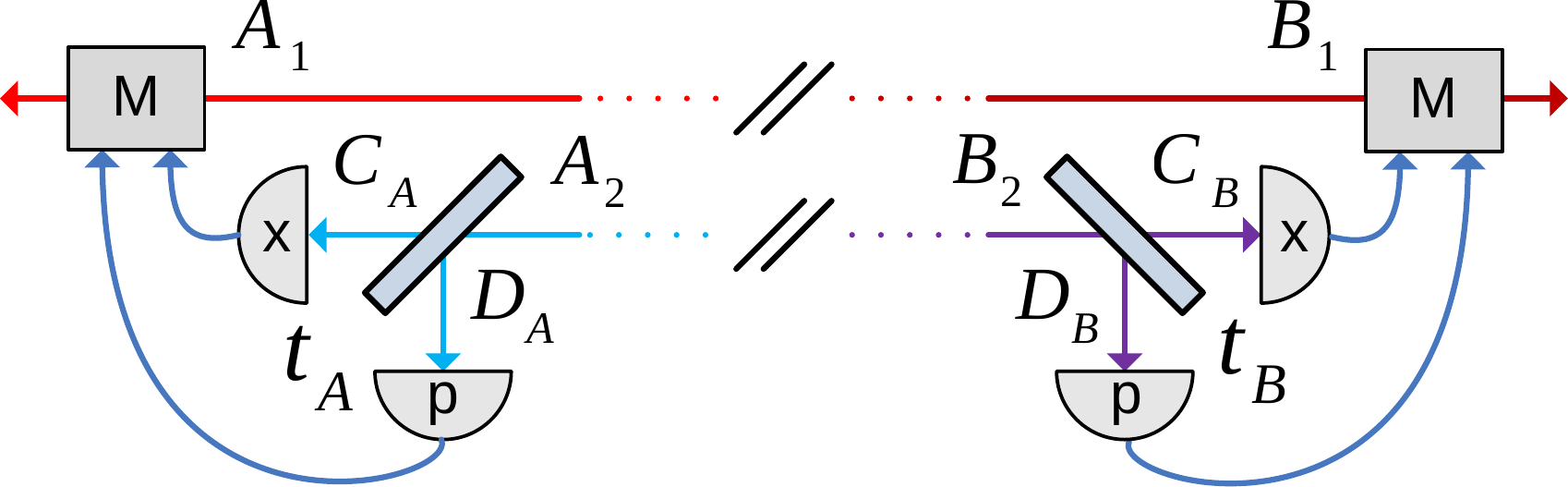}
\caption{Measurement and feed-forward control scheme aimed at compensating cross talk  {and localizing maximum of entanglement } in the pair of modes $A_1,B_1$. The two parties perform generalized Gaussian measurements by splitting modes $A_2,B_2$ on variable beam-splitters $t_A,t_B$ and measuring $x$-quadratures on the modes  $C_A,C_B$ and $p$-quadratures on the auxiliary detection modes   $C_A,C_B,D_A,D_B$. The measurement outcomes are then used  to modulate $A_1,B_1$. The rest of the scheme (source, cross talk and channel) is as in Fig. \ref{scheme_general}. The scheme allows increasing entanglement in modes $A_1,B_1$ at the cost of tracing out modes $A_2,B_2$.}
\label{scheme_ff}
\end{figure}

In order to compensate the negative effect of cross talk  in \textit{only one} entangled pair  {in the scheme given in Fig. \ref{scheme_general}}, we now consider a possibility to apply optimized  Gaussian measurement of one  pair of modes and feed-forward control  of the other  {pair} on the receiving sides and thus to increase the shared entanglement at the cost of losing one of the mode pairs. This method  {relies on the } highly efficient and low-noise homodyne measurement and the high fidelity feed-forward control and coherent displacement \cite{Yokoyama2014,Shiozawa2017}.  Similarly as in the previous Section, we theoretically consider a simple two-mode case  with the feed-forward control applied by Alice and Bob on the first pair of modes $A_1,B_1$ after the measurement on the second pair of modes $A_2,B_2$, as shown in Fig. \ref{scheme_ff}. Instead of a direct interference of multimode signals on a coupler with variable $t_r$, we interfere the signal modes $A_2, B_2$ with the local oscillator  beams in a generalized Gaussian measurement \cite{Miyata2016} on both Alice and Bob sides of the scheme, as shown in Fig. \ref{scheme_ff}. Photocurrents from this detection control the modulation units in the other pair of modes $A_1, B_1$. We therefore assume that both the sender and the receiver perform a general Gaussian measurement \cite{Miyata2016} on the second pair of modes ($A_2,B_2$), i.e., a generalized heterodyne (also known as double-homodyne) measurement, using unbalanced beam-splitters with transmittance values $t_A$ and $t_B$ in Alice's and Bob's sides respectively. The beam-splitters divide each of the measured modes $A_2,B_2$ (into auxiliary detection modes denoted as $ C_A,C_B,D_A,D_B$) and the outputs are then measured in conjugate quadratures using two homodyne detectors (with no loss of generality we assume $x$-quadrature measured on  $C_A,C_B$ and $p$-quadrature measured on $D_A,D_B$). It is advantageous to optimize the measurement shown in Fig. \ref{scheme_ff}, we are therefore searching for optimal $t_A$ and $t_B$ that maximize the logarithmic negativity of the conditional state in the first pair of modes $A_1,B_1$. 
We evaluate conditional matrix of the state in modes $A_1,B_1$ after a homodyne measurement of a quadrature $r$ (being either $x$ or $p$) in mode $K$ (being one of $C_A,C_B,D_A,D_B$). { The covariance matrix of the modes $A_1, B_1$  and $K$ before the measurement is $\gamma_{A_1B_1 K}= \left(
\begin{array}{ccc}
{\gamma}_{A_1} & c_{A_1 B_1} & c_{A_1 K}\\
c_{A_1 B_1}& {\gamma}_{B_1}& c_{B_1 K}\\
c_{A_1 K} & c_{B_1 K} & {\gamma}_K
\end{array}
\right)$. The measurement resulting in an outcome $r_K$ transforms the covariance matrix} as \cite{Eisert2002}
\begin{equation}
\label{eq:cond}
\gamma_{A_1B_1|r_K}=\gamma_{A_1B_1}-\sigma_{A_1B_1,K}\left(R\cdot\gamma_{K}\cdot R\right)^{MP}\sigma_{A_1B_1,K}^T,
\end{equation}
where {$\sigma_{A_1B_1,K}=\left(
\begin{array}{c}
 c_{A_1 K}\\
 c_{B_1 K}
\end{array}\right)$}
 is the correlation matrix between modes $A_1,B_1$ and $K$, matrix R is a diagonal matrix, being either  $\frac{1}{2}(\mathbb{I} +\mathbb{Z} )$ for an $x$-quadrature measurement or   $\frac{1}{2}(\mathbb{I} -\mathbb{Z} )$ for a measurement of $p$-quadrature, and $MP$ stands for Moore-Penrose pseudo-inverse of a matrix, applicable to singular matrices. 
The state in equation (\ref{eq:cond}) can be obtained either by an optimal feed-forward control of the state in modes $A_1,B_1$ or by post-selecting the states in  modes $A_1,B_1$ based on a condition on the measurement outcomes in modes $A_2, B_2$. While the first strategy is deterministic, but requires optimization of feed-forward, the second  one does not rely on optimization, but is probabilistic.   {It only asymptotically approaches the result of feed-forward control, and while the probability of
success greatly reduces while approaching this result  due to reduction of the post-selection interval. On the other hand, it does not require gain optimization and optical modulation as the deterministic method does.} Repeating the conditioning (\ref{eq:cond}) generally for each of four measurements, we arrive at the general conditional matrix $\gamma_{A_1B_1|x_{C_A},x_{C_B},p_{D_A},p_{D_B}}$, which reads $\left(
\begin{array}{cc}
\tilde{\gamma}_{A_1} & \tilde{C}_{A_1B_1}\\
 \tilde{C}_{A_1 B_1}& \tilde{\gamma}_{B_1}
\end{array}
\right)$ with sub-matrices

\begin{footnotesize}
\begin{equation}
\tilde{\gamma}_{A_1}=
\left(
\begin{array}{cc}
\frac{T_2r_B(V-1) [V-t_A (t_c V+t+[V-1])]+V [t_A(V-1)-V]}{t_c T_2r_A r_B \left(V^2-1\right)+[t_A(V-1)-V] [T_2r_B(V-1)+1]} & 0\\
0 & \frac{T_2 t_B(V-1) [t_A(V-1)+1-t_c r_A (V+1)]+V (t_A-t_A V-1)}{t_A(V-1) (T_2 t_B [1+t_c-r_c V]-1)-T_2 t_B(V-1)-1}
\end{array}
\right),
\end{equation}
\begin{equation}
\tilde{\gamma}_{B_1}=
\left(
\begin{array}{cc}
1+T_1(V-1)
-\frac{r_c T_1 r_A \left(V^2-1\right)}{t_A+r_A \left[V-\frac{t_c T_2 r_B \left(V^2-1\right)}{1+T_2 r_B(V-1)}\right]}& 0\\
0 &1+T_1(V-1)-\frac{r_c T_1 t_A \left(V^2-1\right)}{r_A+t_A \left[V-\frac{t T_2 t_B \left(V^2-1\right)}{1+T_2 t_B(V-1)+1}\right]}
\end{array}
\right)
\end{equation}
and
\begin{equation}
\tilde{C}_{A_1B_1}=
\left(
\begin{array}{cc}
\frac{\sqrt{t_c T_1} \sqrt{V^2-1} [(T_2 (1-2 t_A) r_B+t_A )(V-1)-V]}{t_c T_2 r_A r_B \left(V^2-1\right)+[t_A (V-1)-V] [1+T_2 r_B (V-1)]} & 0 \\
0 & \frac{\sqrt{t_c T_1} \sqrt{V^2-1} [1+(t_A (1-2 T_2 t_B)+T_2 t_B )(V-1)]}{t_A (V-1) [T_2 t_B (1+t_c+r_c V)-1]-T_2 t_B (V-1)-1}
\end{array}
\right),
\end{equation}
\end{footnotesize}
where  $r_A \equiv 1-t_A$, $r_B \equiv 1-t_B$. From this two-mode matrix we can evaluate Gaussian entanglement of the conditioned state in terms of logarithmic negativity, which we do numerically. 

First, we observe that the best possible result for any given channel transmittance values and any cross talk level is achieved by using homodyne measurement on the receiver side. It corresponds to $t_B=1$ or $t_B=0$, i.e. measuring either only $C_B$ (homodyne measurement of $x$-quadrature) or only $D_B$ (homodyne measurement of $p$-quadrature) mode in Fig. \ref{scheme_ff}. We further assume  with no loss of generality that $t_B=1$,   which is equivalent to Bob measuring $x$-quadrature of the mode $A_2$ with a homodyne detector. On the other hand, an optimal $t_A$ on the receiver side generally depends on the TMSV state variance $V$, channel parameters (transmittance values and excess noise) and the cross talk coupling $t_c$. In the general case $t_A$ can be found only numerically, but in the limit of large $V$ and noiseless channels, we can find the $t_A$ that maximizes the logarithmic negativity in $\gamma_{A_1B_1|x_{C_A},x_{B_2},p_{D_A}}$ analytically. Then  for $V\to \infty$ and $\varepsilon=0$  {in the limit of the weak cross talk  $t_c\lesssim 1 $ } the optimal $t_A$ on the sender side is independent of the TMSV variance and {can be approximated as}
\begin{equation}
\label{eq:tm_opt}
t_{A_{opt}}= \max \Big[0,1-\frac{(1+T_1) (1-T_2)}{2+2 T_1 (1-T_2)-4 T_2}+O(1-t_c)\Big]
\end{equation}

In the case, when additionally losses are low ($T_1 \to 1, T_2 \to 1$), the optimal setting at the sender side is $t_A=0$. This means that for  {the} highly transmitting channels or low cross talk the best strategy for the feed-forward method of cross talk compensation is to use homodyne measurement on Alice's side and measure complementary quadrature to the one that the receiver (Bob) measures. This phenomena can be explained using continuous-variable quantum erasing, where Alice by measurement prepares a squeezed state in front of the cross talk which, using homodyne measurement of antisqueezed variable and feed-forward control, can be used to eliminate beam-splitter type of cross talk \cite{Filip2003,Andersen2004}. The covariance matrix for a noiseless channel then turns to
\begin{equation}
\gamma_{A_1B_1|p_{A_2},x_{B_2}}=
\left(
\begin{array}{cccc}
 V-\frac{r_c T_2 \left(V^2-1\right)}{1+T_2(V-1)} & 0 & \sqrt{t_c T_1(V^2-1)}  & 0 \\
 0 &  V & 0 & -\sqrt{t_c T_1(V^2-1)}  \\
\sqrt{t_c T_1(V^2-1)}  & 0 & 1+T_1(V-1)& 0 \\
 0 & -\sqrt{t_c T_1(V^2-1)}  & 0 & \frac{V+T_1(V-1) (V t_c+t_c-1)}{V}  \\
\end{array}
\right)
\end{equation}
In a simple case of a perfect channel ($T_i=1$ and $\varepsilon=0$), the logarithmic negativity of the state in modes $A_1,B_1$ (conditioned on homodyne measurement outcomes on both sides) can be expressed analytically as

\begin{equation}
LN_{hom} = - \log_2  \left[\sqrt{1+ t_c \left(V^2-1\right)}-\sqrt{t_c \left(V^2-1\right)}\right].
\label{eq:lnhomT=0}
\end{equation}
It is always larger than the logarithmic negativity in the same pair of  modes  after the crosstalk and in the same perfect channels, but without conditioning, which reads  
 $LN = - \log_2  \left[V-\sqrt{t_c \left(V^2-1\right)}\right]$. The latter equation reaches maximum when $V_{opt}=\frac{1}{\sqrt {1-t_c}}$, while (\ref{eq:lnhomT=0}) is a constantly growing function of $V$. It illustrates the fact that the conditional measurement overcomes the limit on the initial entanglement given by (\ref{eq:vmax}) and generally any need to optimize the initial state. Therefore it allows to use as high initial entanglement as it is experimentally achievable.
When the state is conditioned on the optimal homodyne measurement in a highly transmitting channel, the logarithmic negativity of the state in modes $A_1,B_1$ is not bounded  and in the limit of arbitrarily strong TMSV variance $V \to \infty$ , assuming a noiseless channel ($\varepsilon=0$) tends to
\begin{equation}
\lim_{V\to\infty} LN_{{hom}} = - \frac{1}{2}\log_2 \left[\frac{(1-T_1) [T_1 (1-t_c-T_2)+t_c T_2]}{t_c (1+T_1)^2 T_2}\right]
\label{eq:limhom}
\end{equation}
which may eventually  for large initial variance turn to zero  if the condition $t_c< \frac{(1-T_1) (1-T_2)}{1-T_1 (1-T_2)+3 T_2}$ is met. However, for  realistic values of  relatively weak cross talk ($t_c>0.5$), it happens only for very low $T_2 $. i.e. in channels with high loss. For such channels the homodyne measurement on the sender side is not the optimal one. As the losses in the channel increase, the optimal measurement gets closer to the balanced heterodyne, i.e., to $t_A=1/2$. Moreover, in the most practical cases, i.e. when $T_1$ is close to $T_2$, and the cross talk is small, finding the optimal $t_A$ gives little improvement. In this case it is enough for sender to use homodyning for the highly transmitting (short) channels and heterodyning for the highly attenuating (long) ones, while receiver shall use homodyning in both cases. This result contrasts with the previous results \cite{Sabuncu2010} for recovering quantum information by conditional measurement of the modes leaked into environment, where heterodyne measurement on the receiver side is optimal.

In the case, when the channels are strongly attenuating ($T_{1,2}\to 0$), the optimal $t_A$ given by (\ref{eq:tm_opt}) turns to $1/2$, i.e., the balanced heterodyne detection on the sender side becomes the optimal measurement. In this case, when we also chose the optimal $t_B=1$, the conditional covariance matrix of the state in modes $A_1,B_1$  becomes (assuming no excess noise, $\varepsilon=0$)
\begin{equation}
\gamma_{A_1B_1|x_{A_2},p_{D_A},x_{B_2}}=\left(
\begin{array}{cccc}
\frac{V-T_2(V-1)r_c}{1- T_2 r_c(V-1)}  & 0 & \sqrt{t_c T_1(V^2-1)} & 0 \\
 0 & V & 0 & \frac{-\sqrt{t_c T_1(V^2-1)}}{1- T_2 r_c(V-1)} \\
 \sqrt{t_c T_1(V^2-1)} & 0 &\frac{1+ T_2(V-1)+t_c (T_1-T_2)(V-1)}{1- T_2 r_c(V-1)} & 0 \\
 0 & \frac{-\sqrt{t_c T_1(V^2-1)}}{1- T_2 r_c(V-1)} & 0 &  1+t_c T_1(V-1) \\
\end{array}
\right)
\end{equation} 
The logarithmic negativity is  then as well not bounded and the optimization of the initial entanglement  (\ref{eq:vopt}) is not needed. In the limit of  very strong initial entanglement $V \rightarrow \infty$ (assuming no excess noise, $\varepsilon=0$)  it tends to
\begin{equation}
\lim_{V\to\infty} LN_{{het}} = - \frac{1}{2}\log_2 \left[\frac{(1-t_c T_1) [1-t_c (T_1-T_2)-T_2]}{(1+t_c T_1)^2-(1-t_c) T_2 (1-t_c T_1)}\right]
\label{eq:limhet}
\end{equation}
which is always positive, contrary to the logarithmic negativity before conditioning, which vanishes at (\ref{eq:vmax}).  Eq. (\ref{eq:limhom}) and (\ref{eq:limhet})  show that by optimized conditional measurement one can beat the limit on the maximal initial entanglement given by (\ref{eq:vmax}) and recover the entanglement degraded or destroyed by the cross talk,  although such strong initial squeezing and/or cross talk are not very likely in practical applications.

\begin{figure}[htbp]
\begin{minipage}{0.5 \linewidth}
\includegraphics[width=0.98\linewidth]{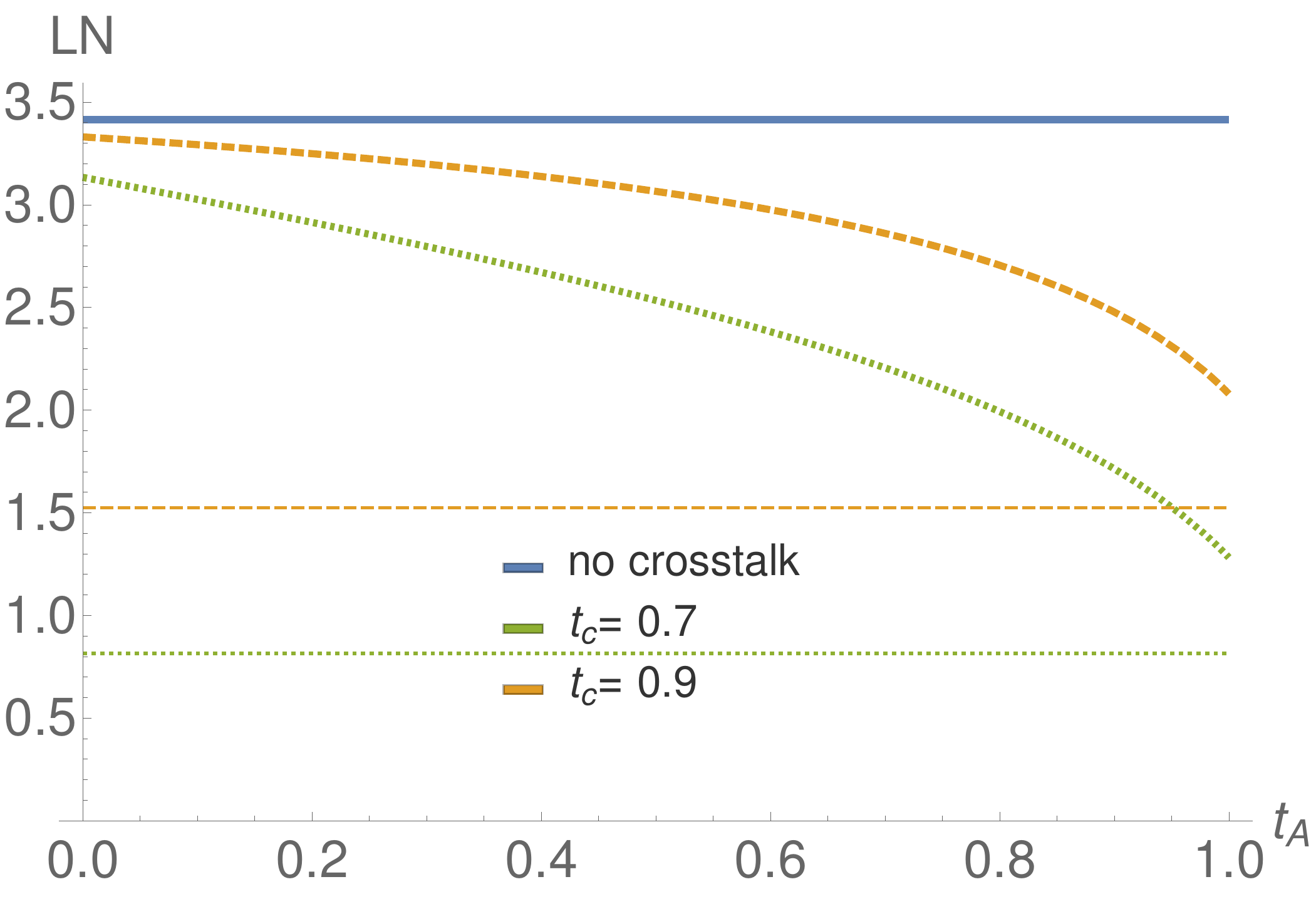}
\end{minipage}
\hfill
\begin{minipage}{0.5 \linewidth}
\includegraphics[width=0.98\linewidth]{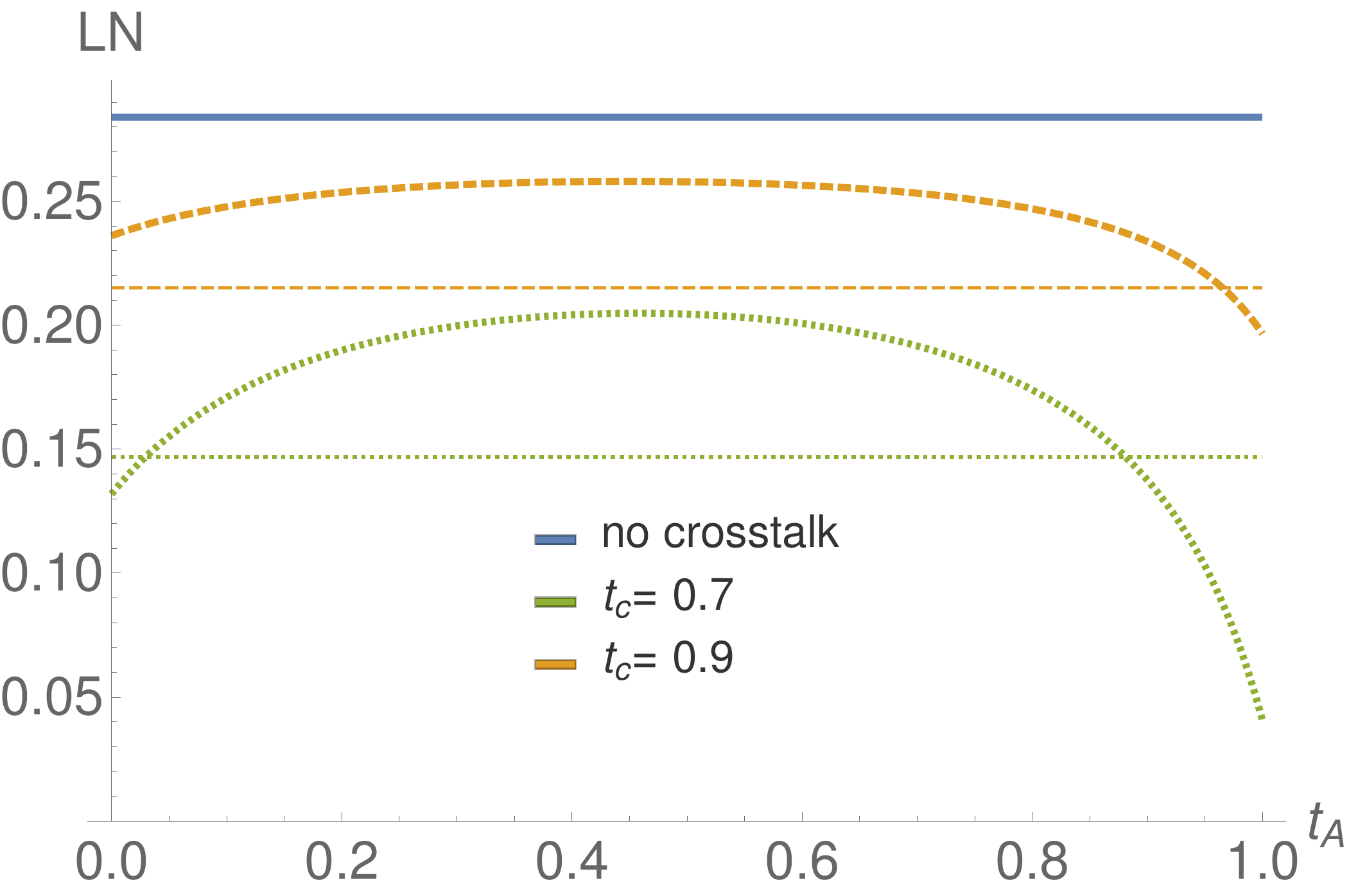}
\end{minipage}
\caption{Logarithmic negativity in the $A_1,B_1$ pair of modes after applying measurement of the pair $A_2,B_2$ and feed-forward control, versus detection setting $t_A$ on the sender side. Left: low unbalanced channel loss ($T_1=-0.4 dB$, $T_2=-0.5 dB$), right: strong unbalanced channel loss ($T_1=-10 dB$, $T_2=-10.5 dB$). On both plots no excess noise is present ($\varepsilon=0$). The thin dashed lines represent the logarithmic negativity in the $A_1,B_1$ pair without the optimized measurement  and feed-forward control, but with the initial state variance being optimized as in (\ref{eq:vopt}).}
\label{ff_stab} 
\end{figure}
In  the general case of intermediate channel attenuation and level of cross talk an optimized heterodyne measurement on the sender side should be chosen by evaluating the respective logarithmic negativities, but entanglement in the remaining pair of modes after the measurement and feed-forward control is always no less, than before these operations. Therefore, optimized measurement and feed-forward can always improve (and fully restore in the case of  almost perfectly balanced channels) entanglement in a single pair of modes at the cost of loosing another pair. We illustrate the efficiency and stability of the method with respect to the sender measurement settings in Fig. \ref{ff_stab} and show that it can largely restore the entanglement in the remaining pair of modes and is stable with respect to the measurement setting $t_A$. As it is evident from Fig. \ref{ff_stab}, any kind of Gaussian measurement on the pair $A_2,B_2$, even not an optimized one, largely restores entanglement in the pair  $A_1,B_1$. The best result is achieved with the homodyne measurement of the $x$-quadrature on the sender side ($t_A=1$) in the low loss channels and with the heterodyne measurement ($t_A=1/2$) in the strong loss channels,  which agrees with approximate analytical results given above in (\ref{eq:tm_opt}).  Remarkably, the optimized measurement allows to recover entanglement that was completely destroyed by  {the} cross talk.

\begin{figure}[htbp]
\begin{minipage}{0.5 \linewidth}
\includegraphics[width=0.98\linewidth]{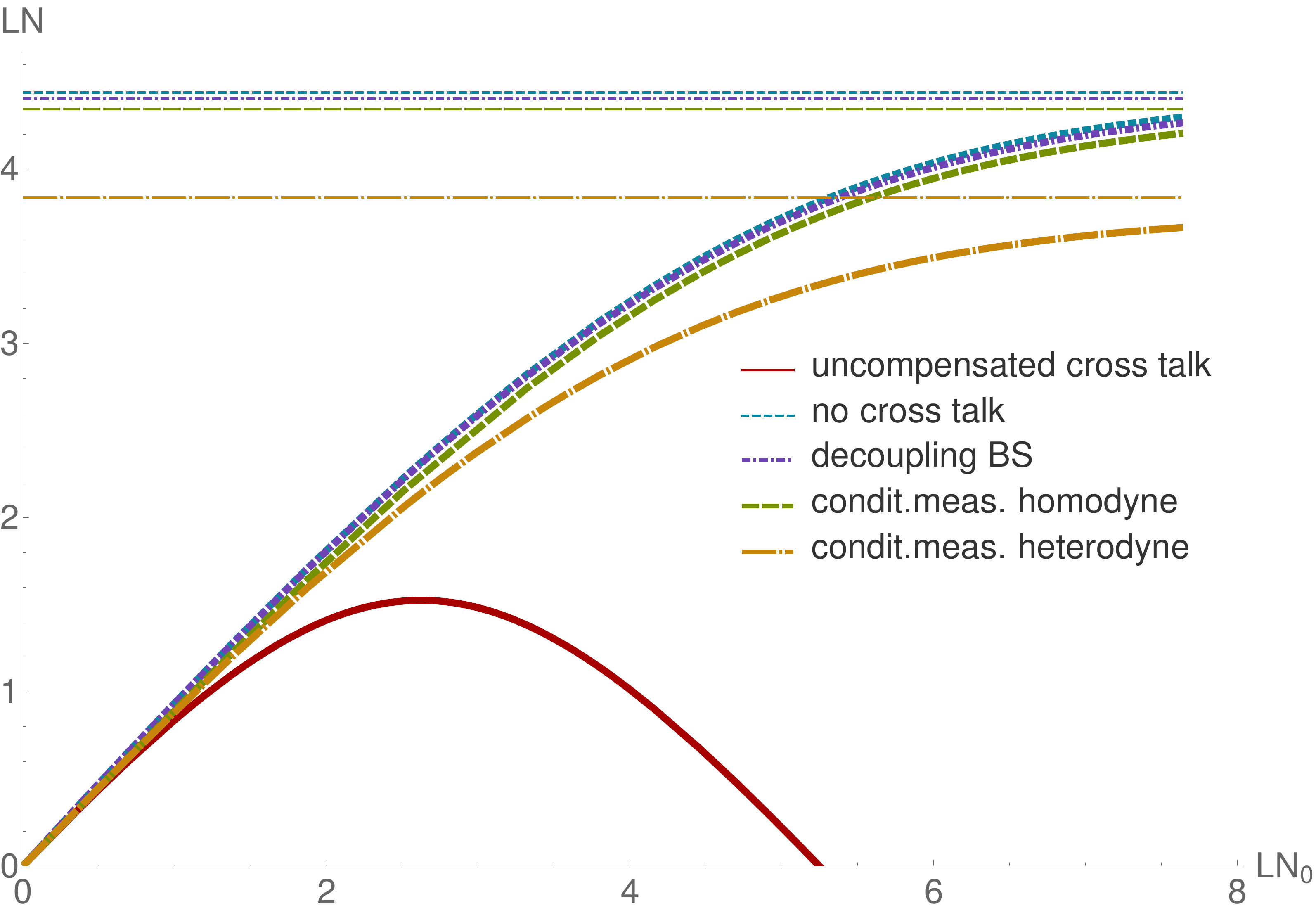}
\end{minipage}
\hfill
\begin{minipage}{0.5 \linewidth}
\includegraphics[width=0.98\linewidth]{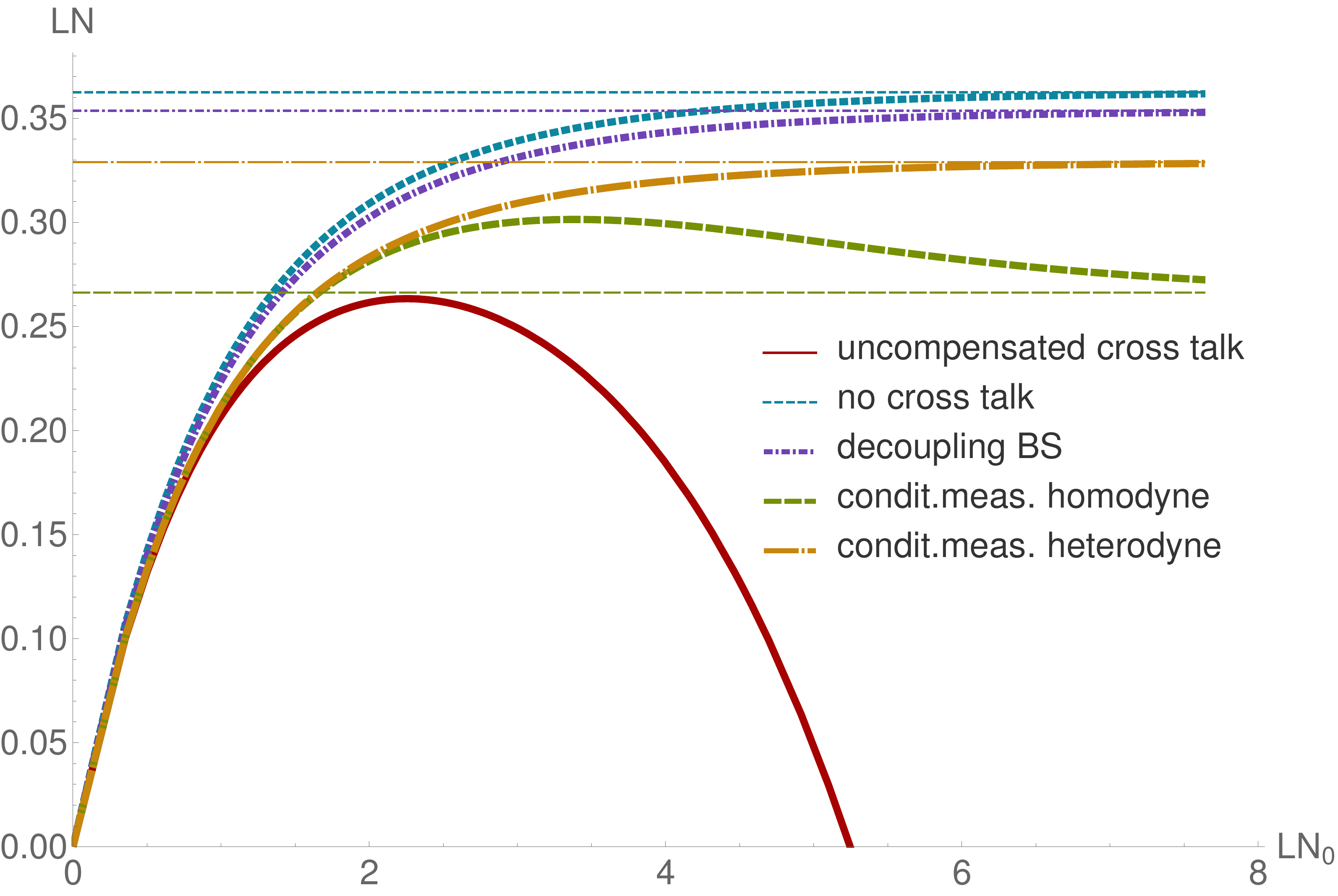}
\end{minipage}
\caption{Logarithmic negativity in the pair $A_1,B_1$ after applying optical interference method or measurement of the pair $A_2,B_2$ and feed-forward control, as indicated in the plots. Cross talk is  $t_c=0.9$ ,  {the optimal} decoupling beam-splitter transmittance $t_r$ is given by eq. (\ref{trInf}), no excess noise ($\varepsilon=0$). Left: low loss unbalanced channels ($T_1=-0.4 dB$, $T_2=-0.5 dB$), right: high loss unbalanced channels ($T_1= -9 dB$, $T_2= -10 dB$). The thin horizontal lines represent the asymptotes for $LN_0\to \infty$ given by  (\ref{eq:ln_decoupl(inf)}) for the decoupling method, (\ref{eq:limhom}) and (\ref{eq:limhet}) for the homodyne and heterodyne measurement method, and (\ref{eq:lnInfty}) for the no cross talk case.}
\label{rev+cond} 
\end{figure}

We also compare the method of optimized measurement and feed-forward to the method of optical interference, suggested in the previous section, as shown in Fig. \ref{rev+cond} for low and high channel losses.

\begin{figure}[htb]
\centering\includegraphics[width=0.75\linewidth]{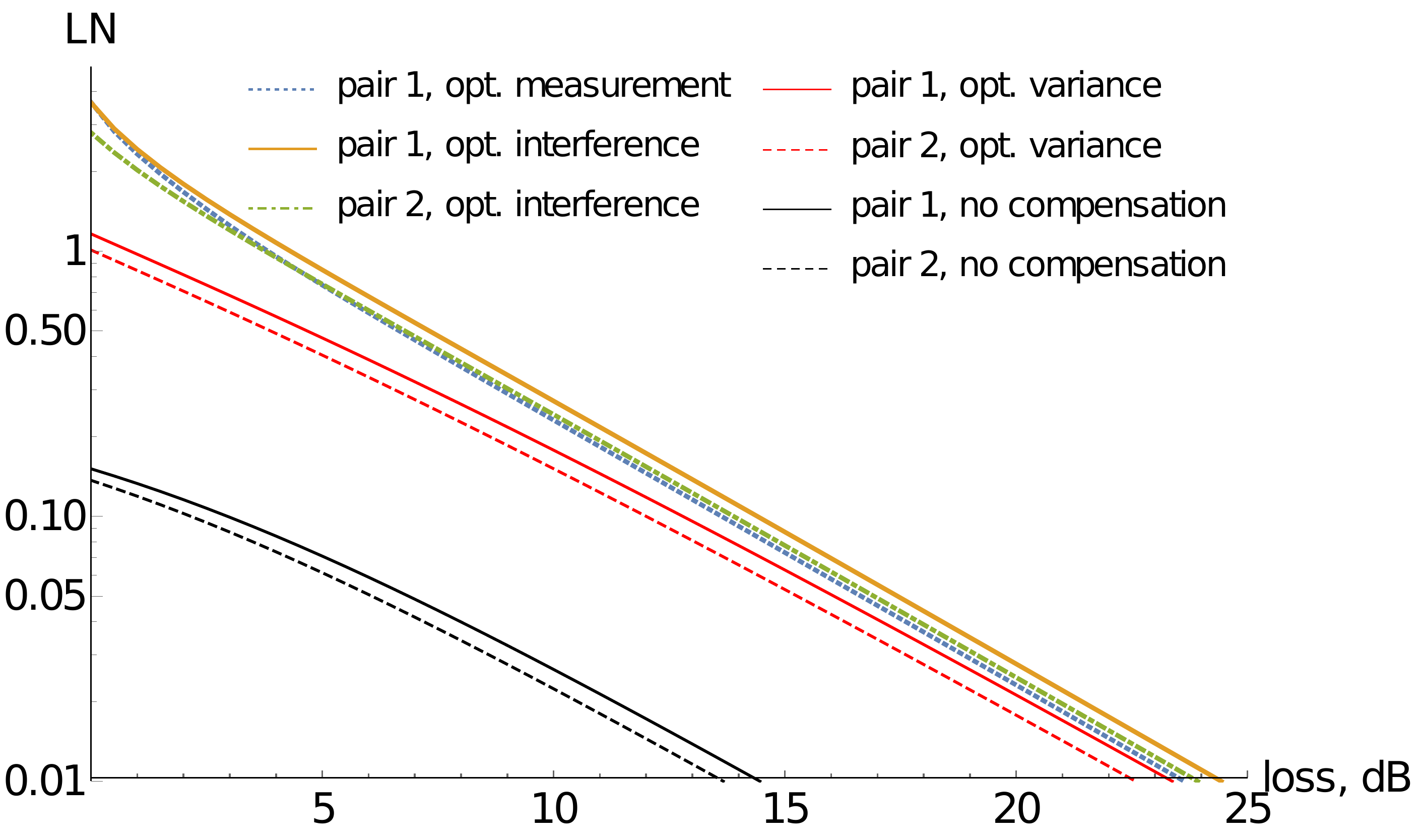}
\caption{Comparison of the methods  {for the cross talk compensation}:  optical interference using a decoupling beam-splitter, and entanglement concentration by optimized conditional measurement and feed-forward control. Plot shows the logarithmic negativity in a pair of modes after each respective method is applied. Initial entanglement is fixed $LN_0= 4.0$, cross talk is $t_c= 0.8$, and transmittances ratio is $T_1/T_2= 1.2$, parameters $t_r$ and $t_A$ are optimized. The ideal case without any cross talk is not shown, but would be indistinguishable from optimized interference for  the given parameters.}
\label{comp} 
\end{figure}

It is evident from the plots, that the method of optimized optical interference is always more efficient in restoring the Gaussian entanglement and, besides, it preserves the multimode structure. Our further analysis shows, that this superiority holds always, unless the channels are strongly unbalanced (with either of the channels being strongly attenuating while another one being almost lossless, $T_i \to 1$ and $T_j \to 0$), which is, however, very unlikely in practical situations. While being less efficient, the measurement strategy can be sometimes nearly optimal and become very close to the optimized measurement strategy, however, the measurement inevitably destroys one pair  {modes}. Both of the methods  increase the amount of entanglement, that can be transferred, and remove the necessity to optimize the initial entanglement.  The lines corresponding to interference method (purple) and to the conditional measurement (green and yellow) continue to the right approaching the asymptotes given by (\ref{eq:ln_decoupl(inf)}), (\ref{eq:limhom}), (\ref{eq:limhet}), and the lines corresponding to the ideal case without cross talk (blue) approach the asymptote (\ref{eq:lnInfty}). We also compare the methods in Fig. \ref{comp} for the same initial entanglement in the large region of transmittance values and in the presence of unbalancing. 
 In Fig.\ref{comp} we plot the Gaussian entanglement for the both pairs of modes in case of the optimized optical interference and show that method allows to restore the entanglement in both the modes contrary to the method of conditional measurement. For a weaker unbalancing the entanglement of both pairs of modes after the optimized interference would overlap. Only one (blue) line in Fig.\ref{comp} corresponds to the conditional measurement method, it demonstrates that although this method is quite effective for restoring entanglement in one pair, it also destroys the second pair of modes.
We have not presented the case with no cross talk on the plot because for the given set of parameters it will be almost fully overlapping with the results for the optimized optical interference method. It illustrates the fact that even for relatively strong cross talk, the optimized interference approach allows to recover  the entanglement almost completely. Both active methods give substantial gain compared to initial state before recovery.
For any cross talk $t_c$, state variance $V$, and channel transmittance $T_1$, $T_2$ both the proposed methods outperform the simple optimization of the initial entanglement shared (or equivalently the variance $V$) as proposed in the Sec. 2. 

Conditional measurement with feed-forward control gives guaranteed entanglement gain no matter how strong is the cross talk and does not rely on our knowledge of cross talk coupling $t_c$,  while discarding one of the modes.  However, this benefit is heavily paid for by a low probability of success of this asymptotic method. On the other hand, the optical interference scheme allows to preserve all the modes. It gives better results than the entanglement concentration with the conditional measurement scheme for  most realistic values of $t_c$, especially in channels with higher loss. In the case of the state having only several modes, both  {of the} methods perform comparably. For multimode states with higher number of modes the advantage of the optimal interference method will be more pronounced as it preserves all the modes, while the conditional measurement method is based on discarding most of them.

\section{Conclusion}

 Entanglement distribution is a key ingredient of modern quantum technology. We considered the effect of cross talk on entanglement distribution using Gaussian multimode twin-beam states and shown that initial entanglement inserted to a multimode link has to be optimized to reach its maximal transmission already if cross talk is weakly present. We then suggest the method of cross talk compensation based on optimized optical interference using adjustable phase control and linear coupling of modes, and show that the method can completely eliminate the negative effect of cross talk once the channel transmittance is balanced, i.e., is the same for all the modes. The method is still very efficient for small unbalancing between the channel transmittance values for different modes and is stable with respect to the linear coupling setting, but requires knowledge of the cross talk strength. This methodology can be extended to many multiplexed modes, however, in such a case the numerical analysis and optimization are needed  for specific real channels, as the model of the cross talk will be more complicated  and vary in practical situations. We therefore leave the analysis of the real cross talk in the source or channel for future  experimental  tests. As an alternative, we suggest the method  of cross talk compensation for one of the  pairs of modes by optimized Gaussian measurement on another pair and feed-forward control, which does not rely on knowledge of cross talk strength and can restore entanglement in the remaining pair, while reducing the multimode structure, and is also limited in the unbalanced channels. The method of optical interference remains more efficient unless the channels are very strongly unbalanced, which is however not likely in the practical situations, and preserves the amount of entangled modes transmitted through the channel. Our methods can be prospective for realization of multimode quantum communication in the presence of cross talk in the sources and channels. {Note that the described cross talk compensation methods distribute entanglement between the modes, but do not increase it beyond the initial entanglement in the source before the cross talk, hence not contradicting the impossibility of entanglement distillation by local Gaussian unitary operations \cite{Eisert2002, Fiurasek2002}. Our} results demonstrate basic principles in a simple case of only nearest-mode cross talk. The further investigation for larger number of modes in the presence of cross talk will require multi-parameter numerical optimizations and possibly application of modern methods of deep machine learning applicable to quantum optics \cite{Dunjko2018} to find efficient strategy of cross talk compensation. Such numerical optimizations in application to complex multimode cross-talk effects will be the subject of future studies.

\section*{Funding}
 {The research leading to these results has received funding from the H2020 European Programme under Grant Agreement 820466 CIVIQ. OK acknowledges support from Palacky University project IGA-PrF-2021-006. OK and VCU acknowledge support from the project 19-23739S of the Czech Science Foundation.} R.F. acknowledges support by the project CZ 02.1.01/0.0/0.0/16\_026/0008460 of MEYS CR and jointly, national funding from the  MEYS and the funding from European Union's Horizon 2020 (2014-2020) research and innovation framework programme under grant agreement No 731473 (project 8C20002 ShoQC). Project ShoQC has received funding from the QuantERA ERA-NET Cofund in Quantum Technologies implemented within the European Union's Horizon 2020 Programme.


\section*{Disclosures}
The authors declare that there are no conflicts of interest related to this article.

\bibliography{crosstalk_compensation}

\end{document}